\pdfoutput=1

\UseRawInputEncoding
\documentclass[twocolumn,dvipsnames]{aastex631}

\usepackage{array}
\usepackage{multirow}
\usepackage{makecell,tabularx}
\usepackage{amsmath}
\shorttitle{Wavelength-dependent atmospheric effects on the DES-SN5YR photometry}
\shortauthors{Lee and Acevedo et al.}
\graphicspath{{./}{figures/}}

\begin{document}

\reportnum{DES-2022-0740}
\reportnum{FERMILAB-PUB-23-040-PPD}

\title{The Dark Energy Survey Supernova Program: Corrections on photometry due to wavelength-dependent atmospheric effects}

\author[0000-0001-6633-9793]{J.~Lee}
\affiliation{Department of Physics and Astronomy, University of Pennsylvania, Philadelphia, PA 19104, USA}
\author[0000-0002-5389-7961]{M.~Acevedo}
\affiliation{Department of Physics, Duke University, Durham, NC 27708, USA}
\author[0000-0003-2764-7093]{M.~Sako}
\affiliation{Department of Physics and Astronomy, University of Pennsylvania, Philadelphia, PA 19104, USA}
\author{M.~Vincenzi}
\affiliation{Department of Physics, Duke University, Durham, NC 27708, USA}
\author{D.~Brout}
\affiliation{Center for Astrophysics $\vert$ Harvard \& Smithsonian, 60 Garden Street, Cambridge, MA 02138, USA}
\author[0000-0002-8687-0669]{B.~Sanchez}
\affiliation{Department of Physics, Duke University, Durham, NC 27708, USA}
\author{R.~Chen}
\affiliation{Department of Physics, Duke University, Durham, NC 27708, USA}
\author[0000-0002-4213-8783]{T.~M.~Davis}
\affiliation{School of Mathematics and Physics, University of Queensland,  Brisbane, QLD 4072, Australia}
\author{M.~Jarvis}
\affiliation{Department of Physics and Astronomy, University of Pennsylvania, Philadelphia, PA 19104, USA}
\author{D.~Scolnic}
\affiliation{Department of Physics, Duke University, Durham, NC 27708, USA}
\author[0000-0003-1899-9791]{H.~Qu}
\affiliation{Department of Physics and Astronomy, University of Pennsylvania, Philadelphia, PA 19104, USA}
\author[0000-0002-1296-6887]{L.~Galbany}
\affiliation{Institute of Space Sciences (ICE, CSIC), Campus UAB, Carrer de Can Magrans, s/n, E-08193 Barcelona, Spain}
\affiliation{Institut d’Estudis Espacials de Catalunya (IEEC), E-08034 Barcelona, Spain}
\author[0000-0003-3221-0419]{R.~Kessler}
\affiliation{Department of Astronomy and Astrophysics, University of Chicago, Chicago, IL 60637, USA}
\author{J.~Lasker}
\affiliation{7 Department of Physics, Southern Methodist University, 3215 Daniel Avenue, Dallas, TX 75275, USA}
\affiliation{Kavli Institute for Cosmological Physics, University of Chicago, Chicago, IL 60637, USA}
\author[0000-0001-9053-4820]{M.~Sullivan}
\affiliation{School of Physics and Astronomy, University of Southampton,  Southampton, SO17 1BJ, UK}
\author{P.~Wiseman}
\affiliation{School of Physics and Astronomy, University of Southampton,  Southampton, SO17 1BJ, UK}
\author{M.~Aguena}
\affiliation{Laborat\'orio Interinstitucional de e-Astronomia - LIneA, Rua Gal. Jos\'e Cristino 77, Rio de Janeiro, RJ - 20921-400, Brazil}
\author[0000-0002-7069-7857]{S.~Allam}
\affiliation{Fermi National Accelerator Laboratory, P. O. Box 500, Batavia, IL 60510, USA}
\author{O.~Alves}
\affiliation{Department of Physics, University of Michigan, Ann Arbor, MI 48109, USA}
\author{F.~Andrade-Oliveira}
\affiliation{Department of Physics, University of Michigan, Ann Arbor, MI 48109, USA}
\author{E.~Bertin}
\affiliation{CNRS, UMR 7095, Institut d'Astrophysique de Paris, F-75014, Paris, France}
\affiliation{Sorbonne Universit\'es, UPMC Univ Paris 06, UMR 7095, Institut d'Astrophysique de Paris, F-75014, Paris, France}
\author[0000-0002-4900-805X]{S.~Bocquet}
\affiliation{University Observatory, Faculty of Physics, Ludwig-Maximilians-Universit\"at, Scheinerstr. 1, 81679 Munich, Germany}
\author[0000-0002-8458-5047]{D.~Brooks}
\affiliation{Department of Physics \& Astronomy, University College London, Gower Street, London, WC1E 6BT, UK}
\author{D.~L.~Burke}
\affiliation{Kavli Institute for Particle Astrophysics \& Cosmology, P. O. Box 2450, Stanford University, Stanford, CA 94305, USA}
\affiliation{SLAC National Accelerator Laboratory, Menlo Park, CA 94025, USA}
\author[0000-0003-3044-5150]{A.~Carnero~Rosell}
\affiliation{Instituto de Astrofisica de Canarias, E-38205 La Laguna, Tenerife, Spain}
\affiliation{Laborat\'orio Interinstitucional de e-Astronomia - LIneA, Rua Gal. Jos\'e Cristino 77, Rio de Janeiro, RJ - 20921-400, Brazil}
\affiliation{Universidad de La Laguna, Dpto. Astrofísica, E-38206 La Laguna, Tenerife, Spain}
\author[0000-0002-4802-3194]{M.~Carrasco~Kind}
\affiliation{Center for Astrophysical Surveys, National Center for Supercomputing Applications, 1205 West Clark St., Urbana, IL 61801, USA}
\affiliation{Department of Astronomy, University of Illinois at Urbana-Champaign, 1002 W. Green Street, Urbana, IL 61801, USA}
\author{J.~Carretero}
\affiliation{Institut de F\'{\i}sica d'Altes Energies (IFAE), The Barcelona Institute of Science and Technology, Campus UAB, 08193 Bellaterra (Barcelona) Spain}
\author{M.~Costanzi}
\affiliation{Astronomy Unit, Department of Physics, University of Trieste, via Tiepolo 11, I-34131 Trieste, Italy}
\affiliation{INAF-Osservatorio Astronomico di Trieste, via G. B. Tiepolo 11, I-34143 Trieste, Italy}
\affiliation{Institute for Fundamental Physics of the Universe, Via Beirut 2, 34014 Trieste, Italy}
\author{L.~N.~da Costa}
\affiliation{Laborat\'orio Interinstitucional de e-Astronomia - LIneA, Rua Gal. Jos\'e Cristino 77, Rio de Janeiro, RJ - 20921-400, Brazil}
\author{M.~E.~S.~Pereira}
\affiliation{Hamburger Sternwarte, Universit\"{a}t Hamburg, Gojenbergsweg 112, 21029 Hamburg, Germany}
\author[0000-0001-8318-6813]{J.~De~Vicente}
\affiliation{Centro de Investigaciones Energ\'eticas, Medioambientales y Tecnol\'ogicas (CIEMAT), Madrid, Spain}
\author[0000-0002-0466-3288]{S.~Desai}
\affiliation{Department of Physics, IIT Hyderabad, Kandi, Telangana 502285, India}
\author[0000-0002-8357-7467]{H.~T.~Diehl}
\affiliation{Fermi National Accelerator Laboratory, P. O. Box 500, Batavia, IL 60510, USA}
\author{P.~Doel}
\affiliation{Department of Physics \& Astronomy, University College London, Gower Street, London, WC1E 6BT, UK}
\author{S.~Everett}
\affiliation{Jet Propulsion Laboratory, California Institute of Technology, 4800 Oak Grove Dr., Pasadena, CA 91109, USA}
\author{I.~Ferrero}
\affiliation{Institute of Theoretical Astrophysics, University of Oslo. P.O. Box 1029 Blindern, NO-0315 Oslo, Norway}
\author{D.~Friedel}
\affiliation{Center for Astrophysical Surveys, National Center for Supercomputing Applications, 1205 West Clark St., Urbana, IL 61801, USA}
\author[0000-0003-4079-3263]{J.~Frieman}
\affiliation{Fermi National Accelerator Laboratory, P. O. Box 500, Batavia, IL 60510, USA}
\affiliation{Kavli Institute for Cosmological Physics, University of Chicago, Chicago, IL 60637, USA}
\author[0000-0002-9370-8360]{J.~Garc\'ia-Bellido}
\affiliation{Instituto de Fisica Teorica UAM/CSIC, Universidad Autonoma de Madrid, 28049 Madrid, Spain}
\author[0000-0001-6942-2736]{D.~W.~Gerdes}
\affiliation{Department of Astronomy, University of Michigan, Ann Arbor, MI 48109, USA}
\affiliation{Department of Physics, University of Michigan, Ann Arbor, MI 48109, USA}
\author[0000-0003-3270-7644]{D.~Gruen}
\affiliation{University Observatory, Faculty of Physics, Ludwig-Maximilians-Universit\"at, Scheinerstr. 1, 81679 Munich, Germany}
\author{R.~A.~Gruendl}
\affiliation{Center for Astrophysical Surveys, National Center for Supercomputing Applications, 1205 West Clark St., Urbana, IL 61801, USA}
\affiliation{Department of Astronomy, University of Illinois at Urbana-Champaign, 1002 W. Green Street, Urbana, IL 61801, USA}
\author[0000-0003-0825-0517]{G.~Gutierrez}
\affiliation{Fermi National Accelerator Laboratory, P. O. Box 500, Batavia, IL 60510, USA}
\author{S.~R.~Hinton}
\affiliation{School of Mathematics and Physics, University of Queensland,  Brisbane, QLD 4072, Australia}
\author{D.~L.~Hollowood}
\affiliation{Santa Cruz Institute for Particle Physics, Santa Cruz, CA 95064, USA}
\author[0000-0002-6550-2023]{K.~Honscheid}
\affiliation{Center for Cosmology and Astro-Particle Physics, The Ohio State University, Columbus, OH 43210, USA}
\affiliation{Department of Physics, The Ohio State University, Columbus, OH 43210, USA}
\author[0000-0001-5160-4486]{D.~J.~James}
\affiliation{Center for Astrophysics $\vert$ Harvard \& Smithsonian, 60 Garden Street, Cambridge, MA 02138, USA}
\author{S.~Kent}
\affiliation{Fermi National Accelerator Laboratory, P. O. Box 500, Batavia, IL 60510, USA}
\affiliation{Kavli Institute for Cosmological Physics, University of Chicago, Chicago, IL 60637, USA}
\author[0000-0003-0120-0808]{K.~Kuehn}
\affiliation{Australian Astronomical Optics, Macquarie University, North Ryde, NSW 2113, Australia}
\affiliation{Lowell Observatory, 1400 Mars Hill Rd, Flagstaff, AZ 86001, USA}
\author[0000-0003-2511-0946]{N.~Kuropatkin}
\affiliation{Fermi National Accelerator Laboratory, P. O. Box 500, Batavia, IL 60510, USA}
\author[0000-0001-9497-7266]{J. Mena-Fern{\'a}ndez}
\affiliation{Centro de Investigaciones Energ\'eticas, Medioambientales y Tecnol\'ogicas (CIEMAT), Madrid, Spain}
\author[0000-0002-6610-4836]{R.~Miquel}
\affiliation{Instituci\'o Catalana de Recerca i Estudis Avan\c{c}ats, E-08010 Barcelona, Spain}
\affiliation{Institut de F\'{\i}sica d'Altes Energies (IFAE), The Barcelona Institute of Science and Technology, Campus UAB, 08193 Bellaterra (Barcelona) Spain}
\author[0000-0003-2120-1154]{R.~L.~C.~Ogando}
\affiliation{Observat\'orio Nacional, Rua Gal. Jos\'e Cristino 77, Rio de Janeiro, RJ - 20921-400, Brazil}
\author[0000-0002-6011-0530]{A.~Palmese}
\affiliation{Department of Physics, Carnegie Mellon University, Pittsburgh, Pennsylvania 15312, USA}
\author[0000-0001-9186-6042]{A.~Pieres}
\affiliation{Laborat\'orio Interinstitucional de e-Astronomia - LIneA, Rua Gal. Jos\'e Cristino 77, Rio de Janeiro, RJ - 20921-400, Brazil}
\affiliation{Observat\'orio Nacional, Rua Gal. Jos\'e Cristino 77, Rio de Janeiro, RJ - 20921-400, Brazil}
\author[0000-0002-2598-0514]{A.~A.~Plazas~Malag\'on}
\affiliation{Department of Astrophysical Sciences, Princeton University, Peyton Hall, Princeton, NJ 08544, USA}
\author{M.~Raveri}
\affiliation{Department of Physics, University of Genova and INFN, Via Dodecaneso 33, 16146, Genova, Italy}
\author{K.~Reil}
\affiliation{SLAC National Accelerator Laboratory, Menlo Park, CA 94025, USA}
\author{M.~Rodriguez-Monroy}
\affiliation{Centro de Investigaciones Energ\'eticas, Medioambientales y Tecnol\'ogicas (CIEMAT), Madrid, Spain}
\author[0000-0002-9646-8198]{E.~Sanchez}
\affiliation{Centro de Investigaciones Energ\'eticas, Medioambientales y Tecnol\'ogicas (CIEMAT), Madrid, Spain}
\author{V.~Scarpine}
\affiliation{Fermi National Accelerator Laboratory, P. O. Box 500, Batavia, IL 60510, USA}
\author[0000-0002-1831-1953]{I.~Sevilla-Noarbe}
\affiliation{Centro de Investigaciones Energ\'eticas, Medioambientales y Tecnol\'ogicas (CIEMAT), Madrid, Spain}
\author[0000-0002-3321-1432]{M.~Smith}
\affiliation{School of Physics and Astronomy, University of Southampton,  Southampton, SO17 1BJ, UK}
\author[0000-0002-7047-9358]{E.~Suchyta}
\affiliation{Computer Science and Mathematics Division, Oak Ridge National Laboratory, Oak Ridge, TN 37831}
\author[0000-0003-1704-0781]{G.~Tarle}
\affiliation{Department of Physics, University of Michigan, Ann Arbor, MI 48109, USA}
\author[0000-0001-7836-2261]{C.~To}
\affiliation{Center for Cosmology and Astro-Particle Physics, The Ohio State University, Columbus, OH 43210, USA}
\author{N.~Weaverdyck}
\affiliation{Department of Physics, University of Michigan, Ann Arbor, MI 48109, USA}
\affiliation{Lawrence Berkeley National Laboratory, 1 Cyclotron Road, Berkeley, CA 94720, USA}

\collaboration{67}{(DES Collaboration)}

\correspondingauthor{Jaemyoung (Jason) Lee}
\email{astjason@sas.upenn.edu}
\correspondingauthor{Maria Acevedo}
\email{maria.acevedo@duke.edu}



\begin{abstract}

Wavelength-dependent atmospheric effects impact photometric supernova flux measurements for ground-based observations. We present corrections on supernova flux measurements from the Dark Energy Survey Supernova Program's 5YR sample (DES-SN5YR) for differential chromatic refraction (DCR) and wavelength-dependent seeing, and we show their impact on the cosmological parameters $w$ and $\Omega_m$.  We use $g-i$ colors of Type Ia supernovae (SNe Ia) to quantify astrometric offsets caused by DCR and simulate point spread functions (PSFs) using the GalSIM package to predict the shapes of the PSFs with DCR and wavelength-dependent seeing. We calculate the magnitude corrections and apply them to the magnitudes computed by the DES-SN5YR photometric pipeline. We find that for the DES-SN5YR analysis, not accounting for the astrometric offsets and changes in the PSF shape cause an average bias of $+0.2$ mmag and $-0.3$ mmag respectively, with standard deviations of $0.7$ mmag and $2.7$ mmag across all DES observing bands (\textit{griz}) throughout all redshifts. When the DCR and seeing effects are not accounted for, we find that $w$ and $\Omega_m$ are lower by less than $0.004\pm0.02$ and $0.001\pm0.01$ respectively, with $0.02$ and $0.01$ being the $1\sigma$ statistical uncertainties. Although we find that these biases do not limit the constraints of the DES-SN5YR sample, future surveys with much higher statistics, lower systematics, and especially those that observe in the $u$ band will require these corrections as wavelength-dependent atmospheric effects are larger at shorter wavelengths.  We also discuss limitations of our method and how they can be better accounted for in future surveys.

\end{abstract}



\section{Introduction} \label{sec:intro}

Type Ia supernovae (SNe Ia) were used in the groundbreaking discovery of the accelerating expansion of the universe \citep{riess1998, perlmutter1999}. Ever since, SN Ia flux measurements have become increasingly precise, with current measurements at the 1\% level with PanSTARRS/Pantheon \citep{Scolnic2018_PanSTARRS_Pantheon}, Pantheon+ \citep{Scolnic_Pantheon+}, and the Dark Energy Survey \citep{brout2019-y3-photometry, smith2020_overview}. Accurate photometry is of fundamental importance in SN cosmology since cosmological parameters are sensitive to the brightness of SNe. In particular, SNe at different redshifts are observed by different bandpasses as their spectra are redshifted, meaning that flux calibrations must be consistent across the bands. As such, it is important to incorporate accurate flux calibrations into the analysis pipeline and quantify the associated uncertainties. As sample sizes grow and other sources of systematic errors become smaller, calibration uncertainties become more and more important as statistical errors become subdominant. 

In this work, we focus on wavelength-dependent ($\lambda$-dependent from hereon) atmospheric effects like differential chromatic refraction (DCR) and seeing effects, which bias PSF photometry flux measurements. We quantify how much these effects impact the Dark Energy Survey Supernova sample (DES-SN5YR), which is the sample of photometrically selected SNe Ia ($\sim 1600$) used for the final cosmological analysis (Vincenzi et al.) from DES. Chromatic corrections considering  variations of the wavelength dependent transmission function depending on 1) the atmospheric variability (changes in the water vapor content) 
and 2) instrumental variations (depending on the location on the focal plane), have been applied in the DES-SN3YR analysis \citep{lasker2019chrom}, but these are unrelated to the $\lambda$-dependent atmospheric corrections we present here, which do not affect the transmission. 

Other $\lambda$-dependent effects in the telescope optics and detector are subdominant for the DES-SN5YR analysis, as we will elaborate in Section \ref{sec:method_coordinate}. The largest optics effect is due to refraction in the lenses, which shifts sources in different bands by different amounts particularly near the boundaries of the focal plane. Such color-dependent radial displacements (also known as lateral color) are shown to be around $0.050\arcsec$/mag and $0.005\arcsec$/mag in the \textit{g} and \textit{r} bands in $g-i$ magnitudes respectively near the edges of the focal plane as described in \citet{bernstein2017astrometric}. Additional effects such as concentric ring-like features as well as brightening near the edges of a CCD caused by stray electrons in the detector, contamination of the dome flat (an image of the dome screen to measure the relative response of pixels across the CCD array) by stray light, color response patterns, and atmospheric extinction are modeled and removed by the DECam photometric model at the mmag level \citep{bernstein2018photometric}.


$\lambda$-dependent atmospheric effects have been investigated in the context of weak gravitational lensing \citep{plazas2012atmospheric, meyers2015impact}, and it was found that shape measurements can be substantially biased when these effects are not considered. DCR effects on quasar photometry have also been explored by \citet{kaczmarczik2009astrometric}. They describe how DCR and measurements of astrometric offsets can, in fact, be used for improving photometric redshifts of quasars. While these effects have not yet been incorporated into any published work, the DES Y6 weak-lensing analysis will include $\lambda$-dependent effects in the PSF, and weak-lensing analysis in LSST is projected to be affected substantially \citep{carlsten2018wavelength}. 

While cosmology using weak-lensing and SN Ia use different measurements (shapes and flux respectively) as probes, it is a worthwhile endeavor to quantify the impact due to $\lambda$-dependent effects, especially before the onset of next-generation ground-based surveys like the Vera C. Rubin Observatory Legacy Survey of Space and Time (LSST) \citep{ivezic2019lsst}. 

\begin{figure}
    \includegraphics[width=0.48\textwidth]{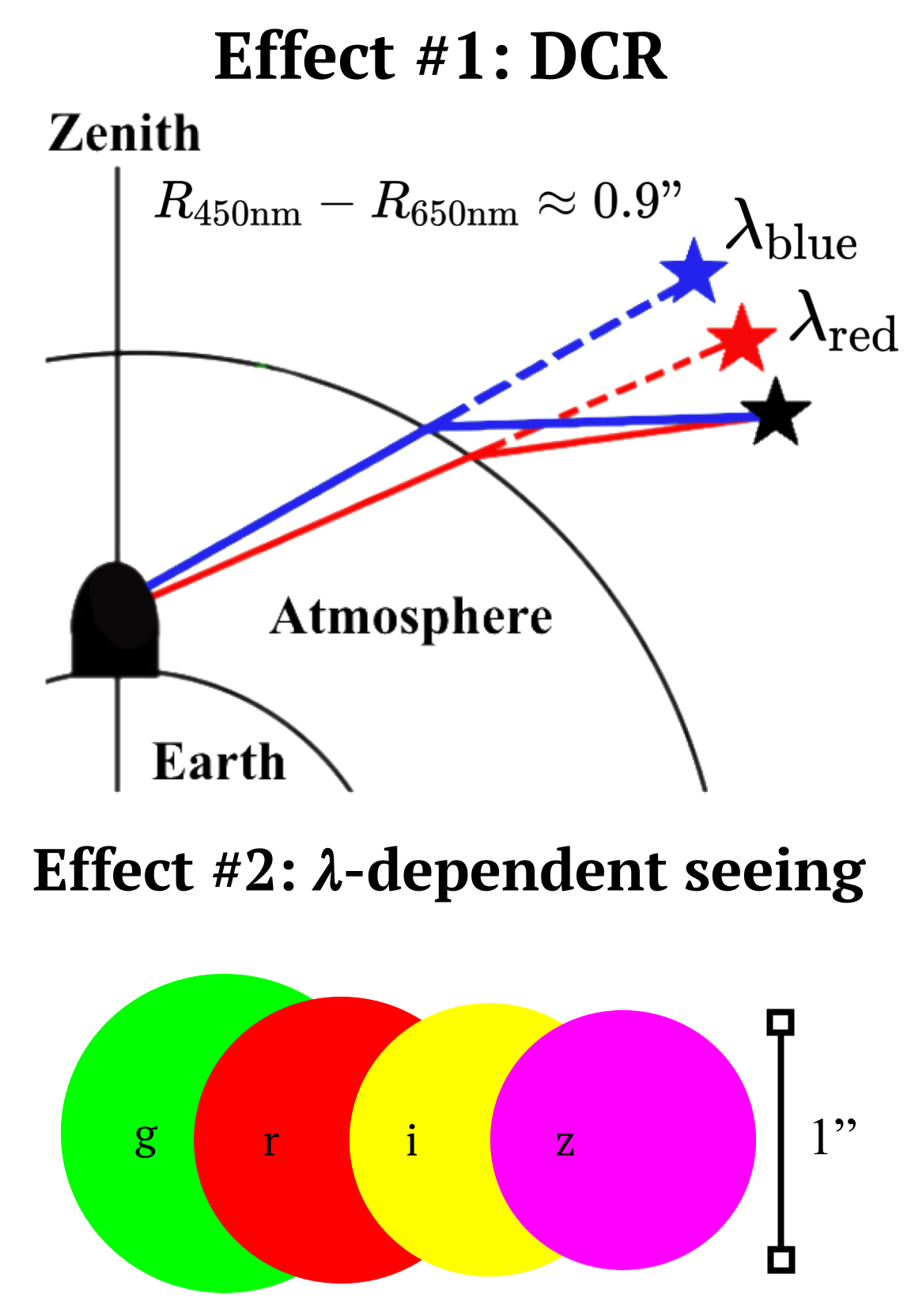}
    \caption
    {Top: A schematic figure showing the effects of DCR as light travels through the Earth's atmosphere; shorter wavelengths are refracted more by the atmosphere, which means that a blue star appears to be higher up in the sky than a red star located at the same position in space. $R_{450\text{nm}}$ and $R_{650\text{nm}}$ denote the amount of refraction by our atmosphere at an AM of 1.4 for 450 nm and 650 nm light respectively.} Bottom: Relative PSF sizes as seen in each of the 4 DES filters (\textit{griz}); PSF sizes increase due to atmospheric turbulence as the wavelength of the filter decreases due to atmospheric effects. 
    \label{fig:refraction}
\end{figure}

We classify $\lambda$-dependent atmospheric effects into two sub-categories: DCR (differential chromatic refraction) and $\lambda$-dependent seeing. 

\begin{figure}
    \includegraphics[width=0.47\textwidth]{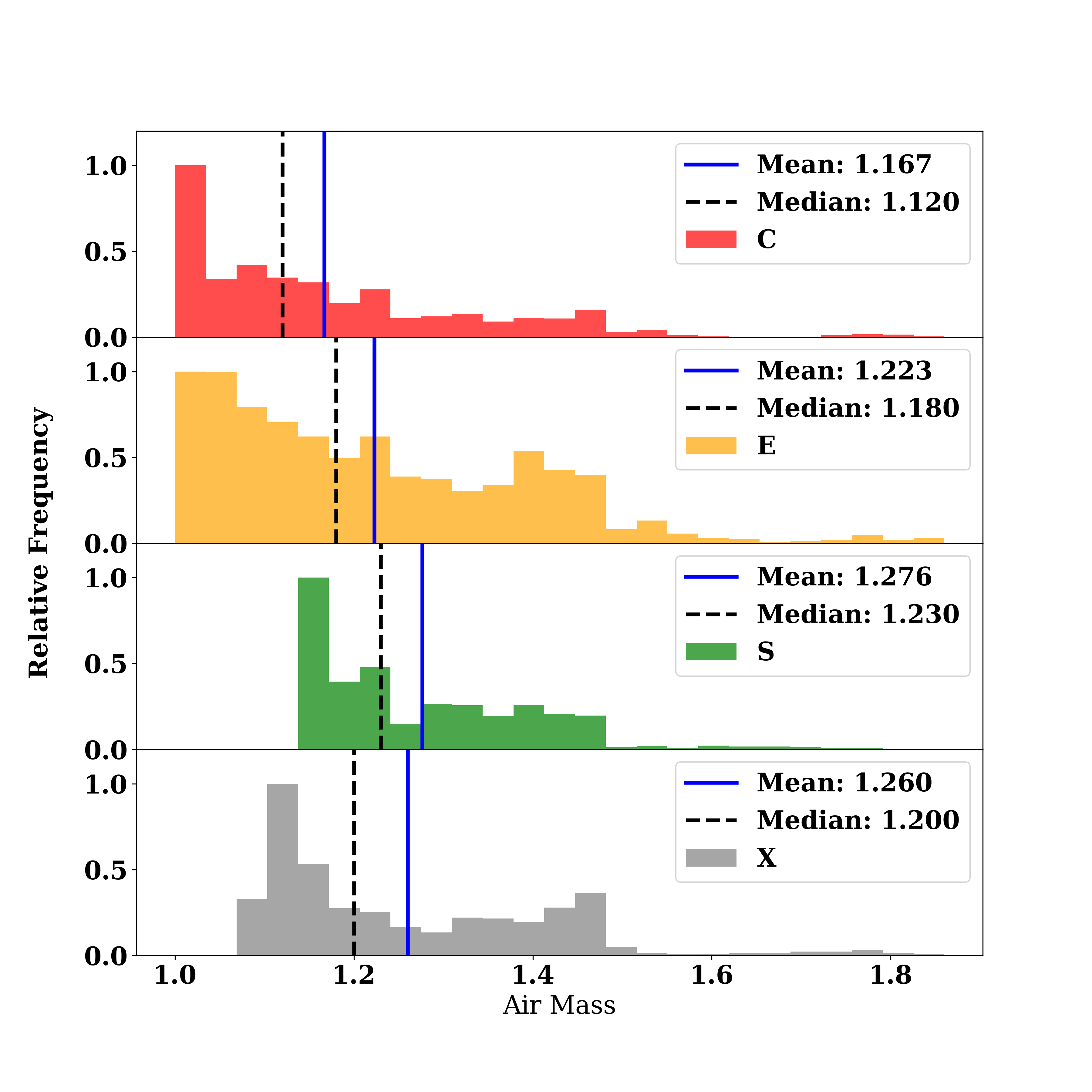}
    \caption
    {Distribution of Air Mass (AM) in the 4 DES SN observation fields (C, E, S, X). The C, E, S and X fields comprise roughly 32.1\%, 22.2\%, 16.3\%, and 29.4\% of the SN observations respectively.} 
    \label{fig:AM_hist}
\end{figure}

\begin{figure}
    \includegraphics[width=0.47\textwidth]{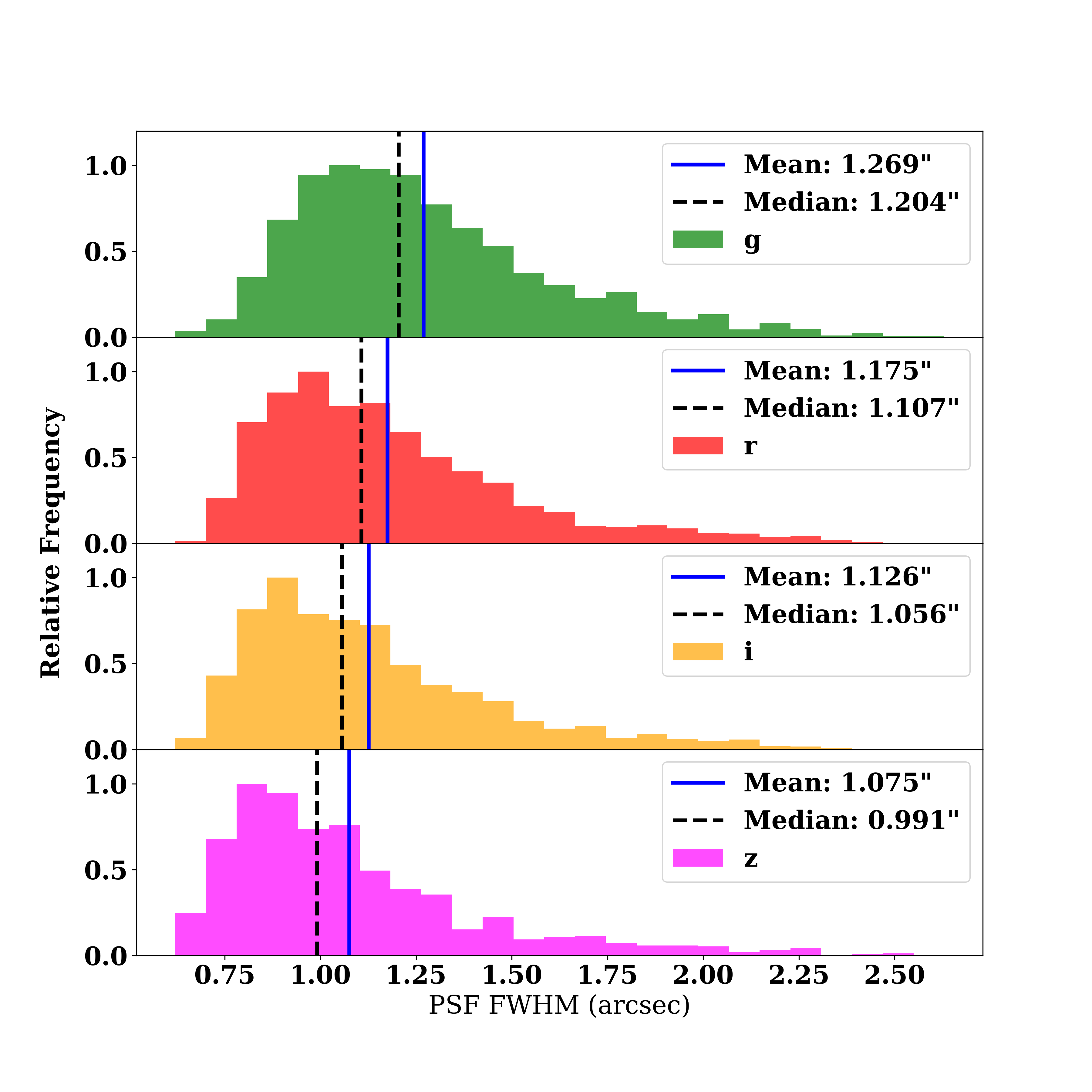}
    \caption
    {Distribution of PSF FWHMs in the 4 DES bands (\textit{griz}).  $\lambda$-dependent seeing effects cause the PSF size to increase at shorter wavelengths.} 
    \label{fig:psf_fwhm_hist}
\end{figure}

\begin{itemize}
\item {\textbf{Effect 1.}} DCR is caused by the index of refraction being $\lambda$-dependent, which is about 1.000281 at 450 nm and 1.000276 at 650 nm \citep{ciddor1996refractive} within our atmosphere (independent of  turbulence), which causes light with relatively shorter wavelengths to be refracted more than longer wavelengths. As shown in the top panel of Figure \ref{fig:refraction}, a blue star would appear higher in the sky to an observer compared to a red star at the same position in space. DCR effects are larger at higher air mass (AM). The AM of a given spot on the sky depends on its coordinates and the hour angle (HA) at which it is observed (or more simply, just its zenith angle). At an AM of about 1.4, light at 450 nm gets refracted about $0.9\arcsec$ more than light at 650 nm by our atmosphere as calculated using Equation 4 of \citet{filippenko1982}. Figure \ref{fig:AM_hist} shows the distribution of AM for each SN Ia observation in the four DES 5YR supernova fields (C, E, S, X) \citep{smith2020_overview}. While most of the observations were taken at $\text{AM} < 1.2$, there are also a substantial number of observations at $\text{AM} > 1.2$.  More details on the AM and HA and the celestial coordinate system are given in Appendix \ref{sec:HA_AM}. Because the SN, host galaxy, and stars that are used to determine the point-spread function (PSF) all have different spectral energy distributions (SEDs), they are all refracted by different amounts in the atmosphere. 

\item \noindent {\textbf{Effect 2.}} $\lambda$-dependent seeing is caused by variations in the atmospheric refractive index due to atmospheric turbulence. The size of the atmospheric convolution kernel (PSF kernel $\theta$) is taken to have a $\theta \propto \lambda ^{\alpha}$ dependency, where $\alpha = -0.2$, predicted by a Kolmogorov turbulence spectrum, is typically used \citep{meyers2015impact}. Note that the PSF size is also a function of air mass ($\theta \propto \text{AM}^{0.6}$). In short, a star would appear larger in a shorter-wavelength filter (\textit{u} or \textit{g}-band) compared to a longer-wavelength filter (\textit{iz}) as shown in the bottom panel of Figure \ref{fig:refraction}. Figure \ref{fig:psf_fwhm_hist} shows the distribution of PSF FWHMs (full-width at half maximum) in the 4 DES bands (\textit{griz}) for the stars observed during the 5 years of DES observations. It is evident that the average PSF sizes increase at shorter wavelength filters. 
\end{itemize}

This paper is as organized as follows. In Section \ref{sec:SMP}, we describe the scene modeling photometry (SMP) \citep{holtzman2008sloan, astier2013photometry, brout2019-y3-photometry} approach used in the DES 5YR analysis and how $\lambda$-dependent effects induce biases in the SN Ia flux measurements using SMP. We discuss the various methods used to calculate the flux bias in Section \ref{sec:methods}. Then, we present the flux corrections needed in the DES 5YR analysis to account for the $\lambda$-dependent effects in Section \ref{sec:DCR_results}. We describe the impact of the flux corrections on cosmological parameters in Section \ref{sec:impact_cosmology}. Lastly, we end with a discussion and conclusion touching on our limitations in Section \ref{sec:conclusion}. 

\section{Scene Modeling Photometry (SMP) and chromatic bias on PSF flux measurements
} \label{sec:SMP}

The scene modeling approach is used in the DES-SN5YR analysis to model a variable transient SN flux and a temporally constant host galaxy \citep{brout2019-y3-photometry}. Here, we briefly outline the scene modeling approach and how DCR and $\lambda$-dependent seeing can cause errors in the flux calculations with SMP. For a more comprehensive review on the SMP approach, see \citet{brout2019-y3-photometry}.

\subsection{Scene Modeling Photometry}

SMP begins with search images processed in the initial stages of the difference-imaging (DiffImg) \citep{kessler2015difference} pipeline. The images are put on a common astrometric solution \citep{2018Morganson_DES_pipeline} and focal plane position-dependent PSFs are created using psfex \citep{bertin2011psfex}. Additionally, the PSFs are constructed from the bright stars on each CCD. Photometric zeropoints are obtained from PSF photometry of tertiary stars and the images are scaled to a common zeropoint. SMP uses an MCMC to simultaneously solve for the static coordinates of the point-like SN and its time-variable fluxes plus the pixelated galaxy model and their uncertainties. 

While the proper motions of stars are considered, SNe are assumed to be ``fixed'' in RA and DEC. The fit is run separately in each band, so the SNe are allowed to have different coordinates in each of the 4 bands. Hence, some of the differences in coordinates across the 4 bands resulting from $\lambda$-dependent effects are accounted for, but $\lambda$-dependent effects within each band across different observations are not. The SN position can be determined at a sub-pixel level. 

 The host galaxy is also assumed to be stationary and modeled to be independent of the AM and atmospheric turbulence.

\subsection{$\lambda$-dependent bias in DES 5YR flux measurements}
\label{sec:dcr_calc}

 Since SMP utilizes PSF photometry, a PSF that is misaligned from the SN or has a different shape from the SN causes a bias in the flux calculation. This is evident in Eq.~\ref{eq:PSF_photometry}, where $F_i$ is the flux of the SN in each pixel $i$, $P_i$ is the value of the PSF in each pixel (we take $\sum_i P_i = 1$), and the supernova flux $F_{\text{SN}}$ is given by

\begin{equation}
    F_{\text{SN}}=\frac{\sum_{i}^{}F_{i}P_{i}}{\sum_{i}^{}P_{i}^{2}} \label{eq:PSF_photometry}
\end{equation}
because $F_\mathrm{SN}$ is a sum of $F_i$, with $P_i$ acting as weights, and even minor shifts in the $P_i$ centroid can bias $F_\mathrm{SN}$.


Hence, PSF flux measurements can be chromatically biased when using SMP in the ways shown in Table \ref{tab:wlt-effects}. 


\begin{table}[h]
\centering
\begin{tabularx}{8.0cm}{|c|p{6.3cm}|} 
\hline
Effects & Description \\ \hline
\multicolumn{2}{|c|}{ DCR effects}\\ \hline
\textbf{A.} & The mismatch between the global SN coordinate and the SN coordinates in each of the observations causes an underestimation of the SN flux. (\textit{COORD})\\ \hline
\textbf{B.} & The shape of the PSF constructed from stars is different from the SN PSF because the average star color is different from the SN. This causes another bias in the SN flux measurement. (\textit{SHAPE-DCR})\\ \hline
\textbf{C.} & The host galaxy position and shape are also shifted by different amounts relative to the SN depending on the AM and its SED. Using an AM independent galaxy model will also potentially cause a bias and additional scatter in the flux measurements. (\textit{CONSTANT GALAXY}, not considered in this paper)\\ \hline
\multicolumn{2}{|c|}{$\lambda$-dependent seeing effects}\\ \hline
\textbf{D.} &  The $\theta \propto \lambda^{-0.2}$ dependency further affects the widths of the reference star PSFs and SN PSFs differently, as they usually have different colors, causing yet another bias. The centroids of the reference star and SN PSFs remain unchanged. (\textit{SHAPE-SEEING})\\ \hline    
\end{tabularx}
\caption{A summary of how $\lambda$-dependent effects impact DES-SN5YR photometry: In the DES-SN5YR analysis, we implement corrections for \textbf{A} (\textit{COORD}) and \textbf{B}+\textbf{D} (\textit{SHAPE}).}
\label{tab:wlt-effects}
\end{table}


For the DES 5YR analysis, we calculate and implement corrections for \textbf{A.} \textit{COORD}, as well as for \textbf{B.} \textit{SHAPE-DCR} and \textbf{D.} \textit{SHAPE-SEEING} (the combined effect of the two denoted as \textit{SHAPE} from hereon) effects. Although the origins of \textbf{B.} \textit{SHAPE-DCR} and \textbf{D.} \textit{SHAPE-SEEING} are different (differential refraction in the atmosphere and atmospheric turbulence respectively), they both result from the PSF shapes being distorted, which is why we chose to combine the two effects into one category. ``$\lambda$-dependent effects" refers to the combination of \textbf{A.} \textbf{B.}, and \textbf{D.} 

Preliminary investigation shows that the \textit{CONSTANT GALAXY} effect does not cause an overall bias on the SN flux measurements, but does cause an additional scatter, which we believe can be attributed to the relative galaxy position and shape causing both flux underpredictions and overpredictions of the SNe. This may explain at least some of the additional scatter seen in SN flux measurements with bright host galaxies (Surface Brightness anomaly) \citep{ brout2019-y3-photometry}, but further analysis of the \textit{CONSTANT GALAXY} effect is beyond the scope of this paper especially since we do not expect an overall bias.

\begin{figure*}
    \centering
    \includegraphics[width=0.98\textwidth]{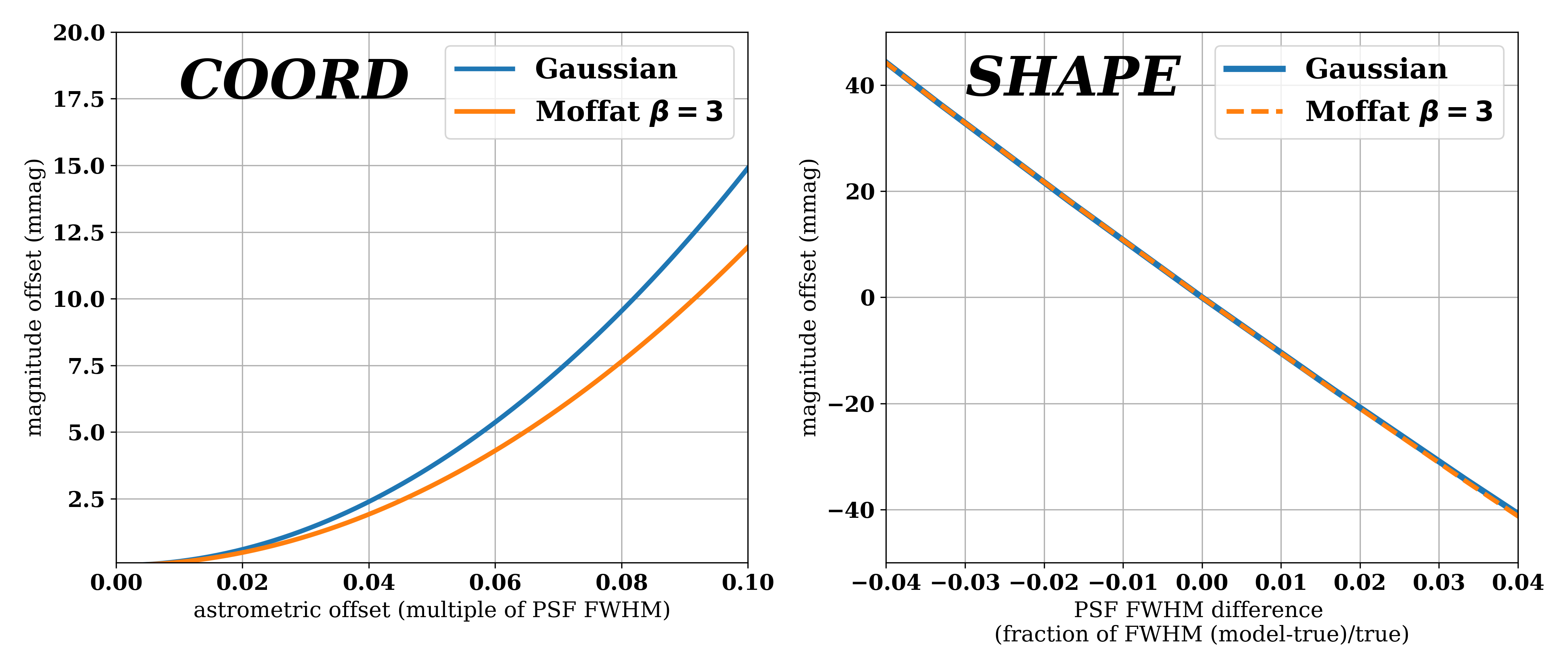}
    \caption{Magnitude offset vs.~astrometric offset in multiples of PSF FWHM, which shows the impact of the \textit{COORD} effect (left) and magnitude offset vs.~PSF size difference in multiples of PSF FWHM, which shows the impact of the \textit{SHAPE} effect (right); an astrometric offset amounting to 10\% of the PSF's FHWM results in about 15 mmag and 12 mmag biases for the Gaussian and Moffat profiles respectively while a 3\%  mis-estimation of the PSF FWHM can result in about 30 mmag biases for both the Gaussian and Moffat.}
    \label{fig:psf_mag_offset}
\end{figure*}

In Figure \ref{fig:psf_mag_offset}, we show the resulting magnitude offset in mmags depending on the astrometric offset for the \textit{COORD} bias and the offset depending on the PSF FWHM difference for the \textit{SHAPE} bias for the Gaussian and Moffat (with $\beta = 3$ as typically assumed) profiles \citep{moffat1969} to illustrate the potential scale of the $\lambda$-dependent effects. The (2D) Gaussian profile is described by $I_{\mathrm{Gaussian}}(r) \propto e^{-r^2/2\sigma^2}$ while the Moffat is described by $I_{\mathrm{Moffat}}(r) \propto \big(1 + (r/r_0)^2\big)^{-\beta}$, where $I(r)$ is the flux, $r$ the distance from the center of the profile, $\sigma$ the width of the 2D Gaussian, and $r_0$ the scale radius of the Moffat. Note when $\beta \xrightarrow[]{} \infty$, the Moffat profile becomes a Gaussian. It can be seen that an astrometric offset that amounts to 10\% of the PSF's FWHM results in about 15 mmag and 12 mmag biases for the Gaussian and Moffat profiles respectively, while an over- or underestimation of the PSF FWHM by 3\% can result in about 30 mmag biases for both the Gaussian and Moffat.

\section{Methods} \label{sec:methods}

In this section, we describe how the \textit{COORD} and \textit{SHAPE} effects are calculated and implemented into the DES 5YR analysis as corrections to the SMP photometry. We use the model $g-i$ magnitudes of the $\sim 1600$ DES candidates with SALT2 \citep{guy2007salt2} fits to the SMP light curves as proxies for the SN SEDs. The fits use forced photometry and the SALT2 model is as trained in \citet{taylor2023salt2}. Although the DES-SN 5YR cosmological analysis is done with SALT3 \citep{kenworthy2021salt3}, the $g-i$ magnitude differences between SALT2 and SALT3 are minimal (less than 0.003 as shown in \citet{taylor2023salt2}), so we do not expect a noticeable difference. Additionally, we need to use the model $g-i$ magnitudes of the SNe without the $\lambda$-dependent effects to calculate the $\lambda$-dependent effects to be precise, but we use the SALT2 fits from observed (and hence including $\lambda$-dependent effects) SMP light curves. The resulting bias is minimal since the $\lambda$-dependent shifts are smaller than the flux uncertainty. We do not provide \textit{COORD} and \textit{SHAPE} corrections for SN if the $g-i$ magnitude is not available, at the high redshifts where the SN is not detected in the \textit{g} band (roughly 20\% of observations). 

For the PSF shapes, we use a Moffat profile, which captures the wings of a PSF better than a Gaussian profile, with $\beta = 3$ before $\lambda$-dependent effects. We confirmed that the Moffat profile is a much better fit to the psfex PSF than the Gaussian by fitting both profiles to the psfex PSF and calculating the $\chi^2$ of the fits. We also found that the Moffat profile describes the PSF accurately enough for our purposes.


\subsection{COORD - A data-based approach} \label{sec:method_coordinate}

\begin{figure}
    \includegraphics[width=0.47\textwidth]{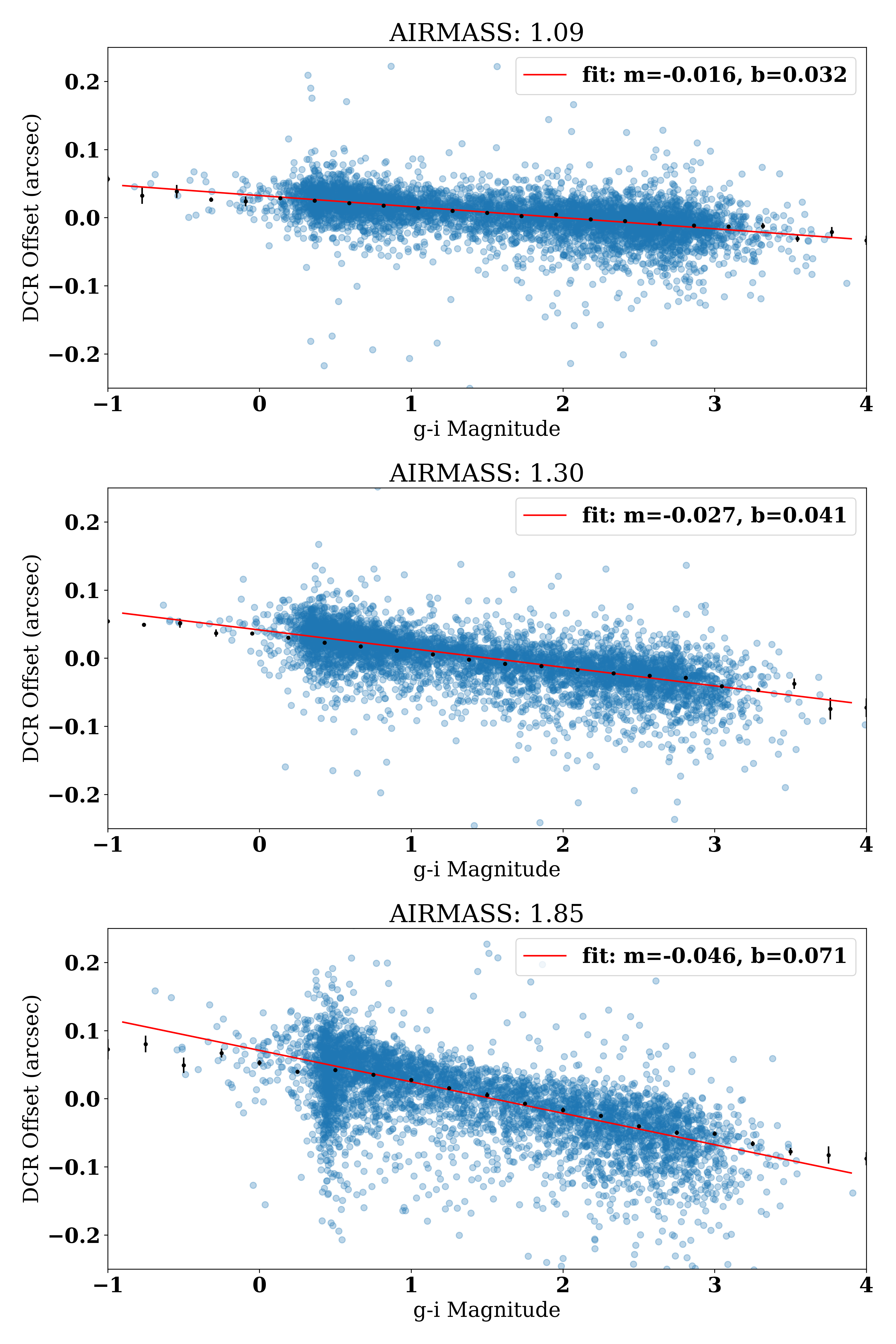}
    \caption
    {Observed astrometric offsets of field stars as a function of their $g-i$ color of three \textit{g}-band exposures taken at different AM values. The black points represent the binned median while the line is the linear best-fit with slope $m$ and y-intercept $b$.     \label{fig:g-i_intercept}} 

\end{figure}

For the \textit{COORD} effect, we determine the relative astrometric offset of the SN from the mean coordinates in each exposure determined by the calibration stars. To compute these relative offsets, we first determine the expected offsets in each exposure by measuring the astrometric offsets of stars on the image relative to the catalog coordinates. These are shown in Figure~\ref{fig:g-i_intercept} for $\text{AM} = 1.09, 1.30, 1.85$. We fit a linear function to the median offset as a function of color per exposure and tabulate the slope and intercept. The magnitude of the slope increases with increasing AM as expected. The $x$-intercept occurs at $g-i = 1.666$, 1.430, 1.520, 1.561 for C, E, S, X fields respectively, which corresponds to the average color of the catalog stars and therefore experience zero astrometric shift.  Bluer (redder) stars are shifted toward (away from) zenith. Our observations show no significant shifts in the azimuthal direction as expected.  A detailed calculation of the astrometric offsets is given in Appendix \ref{sec:trig}.


For each SN in each band, we first calculate the DCR coordinate shifts relative to its fiducial coordinate. Using these shifts, we calculate the mean (global) coordinate of the SN by weighting by the $(\text{S/N})^2$ of the observations. We then compute the relative offsets for each of the observations by subtracting out this mean coordinate. These relative offsets are finally used to predict the corresponding photometric corrections (\textit{COORD}) using the results described in Section~\ref{sec:dcr_calc} and shown in Figure~\ref{fig:psf_mag_offset}. Hence, the photometric correction for each observation depends on the slope (AM) and $x$-intercept of each exposure (Figure~\ref{fig:g-i_intercept}.) 

Although the \textit{COORD} effect can be calculated using the simulation-based approach described in Section~\ref{sec:method_shape}, we chose the data-based approach because we can directly measure the offset of an SN relative to the mean (global) coordinate accurately and choosing a reference star spectrum of the PSF for the \textit{SHAPE} effect requires an approximation as described in Section~\ref{subsubsec:eff_wave}. We calculated the \textit{COORD} offsets with respect to the mean coordinate using both the data-based approach and the simulation-based approach and found that the two methods are consistent.

Color-dependent radial displacements mentioned in Section \ref{sec:intro} result in a similar bias to the \textit{COORD} effect since the SN positions will shift particularly near the edges of the focal plane depending on their $g-i$ colors. DES-SN observations have very small dithers, so each SN is detected at almost exactly the same location on the focal plane. However, since the $g-i$ magnitude of a given SN changes by 1.5 to 2 mag throughout its evolution for $0.0 < z < 1.0$, the radial displacements around the edges can be up to $0.1\arcsec$ in the $g$ band. We quantified this effect and found that the size of this bias is, at worst, similar to the bias caused by positional shifts in the SN due to $\lambda$-dependent effects, but much smaller than the bias caused by shape distortions in the SN due to $\lambda$-dependent effects.

\subsection{SHAPE - A simulation-based approach}
\label{sec:method_shape}
Unlike the data-based method adopted to compute the \textit{COORD} effect corrections, we use image simulations to model the \textit{SHAPE} effect and calculate the amount of flux bias caused by the \textit{SHAPE} effect, since it is not viable to calculate such complicated shape distortions analytically.  
We summarize this procedure below.
\begin{enumerate}
    \item We generate pixelated images (with the pixel size being around $0.263\arcsec$ as with DECam) of the SN and reference star PSF, for various AMs, FWHMs, using GalSIM, which includes both DCR and $\lambda$-dependent seeing effects. The $g-i$ color is varied for the SN, while the reference star's color is fixed so that no DCR shift occurs on average.
    \item We then use the simulated reference star PSF to measure the SN flux and calculate the flux bias depending on the AM, FWHM, and $g-i$  color of the SN. The results are appended to a look-up table which contains four columns: AM, FWHM, $g-i$ color, and the flux bias. 
    \item To calculate the magnitude correction needed for a SN, we take the AM, FWHM, and $g-i$ color of the observed SN, perform a 3D interpolation using the table described above, then convert the flux bias into magnitude corrections. 
\end{enumerate}

For the \textit{SHAPE} effect, we only consider the \textit{g} and \textit{r} bands as we expect the effect to be negligible in \textit{i} and \textit{z}. In the subsequent subsections, we describe the important details in our method.

\subsubsection{GalSIM} \label{subsubsec:galsim}

We use the GalSIM\footnote{\url{https://galsim-developers.github.io/GalSim/\_build/html/index.html}} \citep{rowe2015galsim} python package to predict the fractional change of the PSF FWHM based on the $g-i$ color of the SN compared to the average PSF determined from the stars. We first define a stellar or SN profile, which are taken to be $\delta$-functions (point sources). They are then convolved with a realistic PSF model that includes the atmospheric seeing properties such as DCR and $\lambda$-dependent seeing. We next input World Coordinate System (WCS) information to the image headers. Lastly, PSF photometry is performed using the DES filter response functions. Since GalSIM requires the spectral energy distribution (SED) of a profile to calculate the $\lambda$-dependent effects, we use stellar SEDs of main sequence stars given in \citet{pickles1998stellar}\footnote{\url{https://www.eso.org/sci/facilities/paranal/decommissioned/isaac/tools/lib.html}} as proxies for $g-i$ colors of SN and reference star PSFs. For sanity checks, we use the SN template given in \citet{hsiao2007_sn_template}\footnote{\url{http://astrophysics.physics.fsu.edu/\~hsiao/data/}}. In Section \ref{sec:conclusion}, we discuss the limitations of using $g-i$ colors of stars rather than full SN SEDs.



\begin{figure}
    \includegraphics[width=0.47\textwidth]{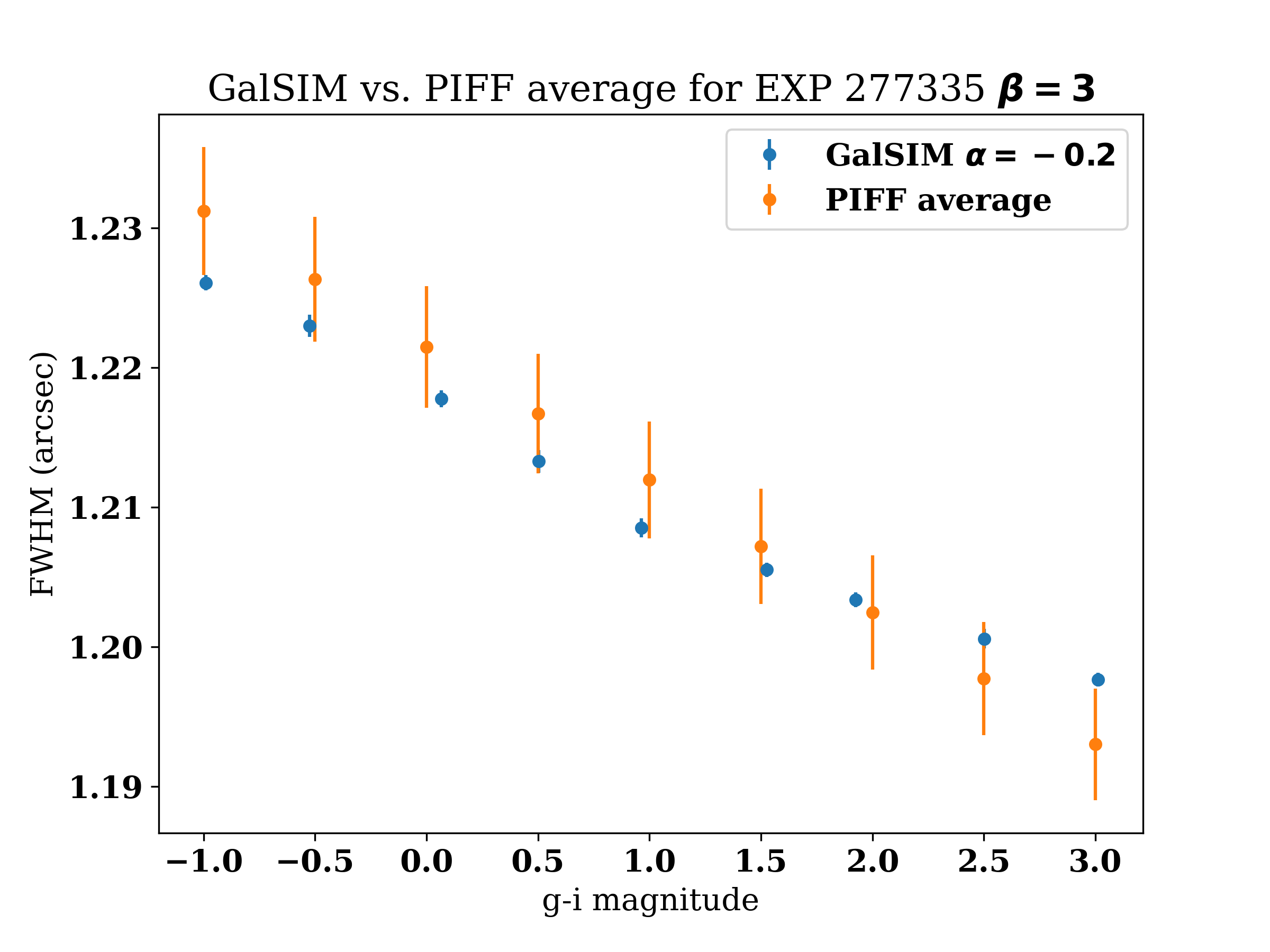}
    \caption
    {PSF FWHM vs.\ $g-i$ magnitude using GalSIM and PIFF for one exposure; the slope varies between different CCDs for PIFF, so the average over all the CCDs are shown here. GalSIM's predictions are comparable to this average slope.     \label{fig:fwhm_PIFF_GalSIM}} 

\end{figure}

Figure \ref{fig:fwhm_PIFF_GalSIM} shows the predicted PSF FWHM values using GalSIM as a function of the $g-i$ magnitude. Bluer $g-i$ magnitudes (smaller) should result in larger FHWMs, as shown in Figure~\ref{fig:psf_fwhm_hist}. The slope of this relation matches the average PSF size measurement from PIFF\footnote{\url{http://rmjarvis.github.io/Piff/\_build/html/index.html}}, which is the state-of-the-art package developed for DES PSF measurements. PIFF's slope varies considerably depending on the CCD, which reflects the telescope's optical properties.  While we do not use PIFF generated PSFs to calculate the \textit{SHAPE} effect, GalSIM's slope is consistent with the PIFF-generated PSFs on average, which validates our approach.


\subsubsection{Effective Wavelength} \label{subsubsec:eff_wave}

To use GalSIM to calculate the DCR shifts and $\lambda$-dependent seeing effects, we choose a reference wavelength and compute DCR effects relative to that wavelength. We do this because the data are treated in a similar way where the initial astrometric solution and PSFs are constructed from field stars that span a wide range in colors.  As such, they are correct for the average-color stars only. To determine the reference wavelength, we choose the star with $g-i$ color that exhibits zero DCR coordinate shift (the $x$ intercept in Figure~\ref{fig:g-i_intercept}) as the reference star.  The average color of stars differs slightly from field to field, as given in Section \ref{sec:method_coordinate}.  We adopt the average $x$ intercept across all SN fields, which occurs at $g-i = 1.54$, corresponding to a K5V star. In practice, PSFs are built using all the stars in each exposure, so assuming a particular value of $g-i$ magnitude is correct on average, but still an approximation. Using this approximation results in a very small bias at less than the 1 mmag level, which is small enough to be ignored. Once a reference star is chosen, we calculate the effective wavelength using the reference star SED for a given filter using the expression,
\begin{equation}
    \lambda_{\text{eff}} = \frac{\int_{0}^{\infty}\lambda S\left(\lambda\right)F\left(\lambda\right)d\lambda}{\int_{0}^{\infty}S\left(\lambda\right)F\left(\lambda\right)d\lambda},
\end{equation}
where $S(\lambda)$ is the flux of the star at a given $\lambda$ while $F(\lambda)$ is the filter function. For the DES \textit{g}-band and \textit{r}-band for which the \textit{SHAPE} corrections are calculated, $\lambda_{\text{eff}}$ are 490.8 nm and 643.2 nm respectively. We then use GalSIM to compute all DCR effects relative to this star.  

\subsubsection{Look-up Table}

To calculate the \textit{SHAPE} corrections, we first use GalSIM to generate SNe and reference star PSFs for a range of AM (1.0 to 2.5), PSF FWHMs ($0.8\arcsec$ to $1.5\arcsec$), and $g-i$ magnitudes (-1.0 to 4.0), which covers the range of observing conditions and colors of SNe in the DES 5YR data.  We then calculate the magnitude offset for each set of parameters by centering the PSF onto the SN and measuring the PSF flux, to consider only the \textit{SHAPE} effect. We next create look-up tables for the \textit{g} and \textit{r} bands with magnitude corrections on a grid of AM, PSF FWHMs, and $g-i$ values.  For each SN observation with a valid $g-i$ magnitude, we then use grid interpolation (\verb|scipy|'s \verb|RegularGridInterpolator|) to obtain the magnitude correction. We note that the PSF FWHM changes due to the \textit{SHAPE} effect and the FWHM values used in the look-up table are acquired by fitting Moffat profiles after the \textit{SHAPE} effect is induced. Although the stellar FWHM values of DES exposures can be as large as $2.5\arcsec$, the \textit{SHAPE} bias decreases as the PSF FHWMs increase and plateaus around $\text{FWHM} = 1.5\arcsec$.


\section{Corrections} \label{sec:DCR_results}

In this section, we show the amount of magnitude corrections (MAGCOR) needed for the DES-SN5YR observations due to $\lambda$-dependent atmospheric effects. The magnitude corrections are added to the original magnitudes ($m_{\mathrm{OLD}}$) calculated without $\lambda$-dependent effects to obtain the corrected magnitudes ($m_{\mathrm{NEW}}$).
\begin{equation}
    m_{\mathrm{NEW}} = m_{\mathrm{OLD}} + \mathrm{MAGCOR}
\end{equation}

\subsection{COORD Corrections}

\begin{figure}
    \centering
    \includegraphics[width=0.47\textwidth]{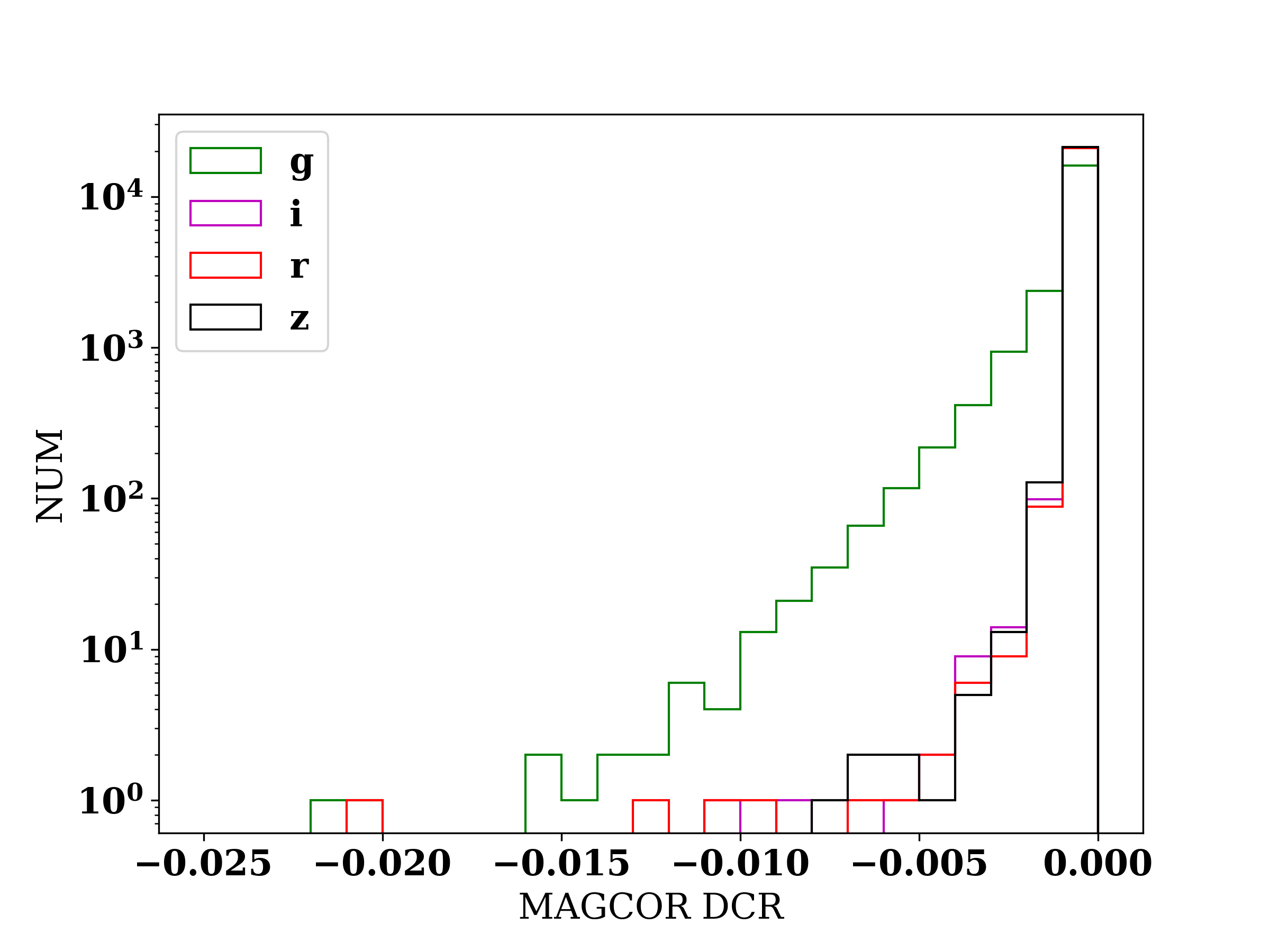}
    \caption{Magnitude corrections histogram for the \textit{COORD} effect; corrections are always negative and about 1.3\% of corrections in the \textit{g} band are larger than 0.005 mag in size. Note that the horizontal axis is in units of mag and the vertical axis is logarithmic.}
    \label{fig:mag_corr_hist_coord}
\end{figure}

For the \textit{COORD} effect, we provide magnitude corrections for  \textit{griz} bands for SN candidates with valid \textit{g} and \textit{i} magnitudes. In Figure \ref{fig:mag_corr_hist_coord}, we see that the magnitude corrections are always negative as the flux is always underestimated. 

The corrections are, on average, larger for the \textit{g}-band as it has the shortest wavelength. Taking the absolute value of these corrections, approximately 0.1\% of the \textit{g} band \textit{COORD} corrections are larger in magnitude than 0.01 mag, while 1.3\% are larger than 0.005 mag. For the \textit{r}, \textit{i}, and \textit{z} bands, this effect is smaller with less than 0.03\% greater than 0.005 mag and less than 0.01\% greater than 0.01 mag.

\subsection{SHAPE Corrections}

\begin{figure}
    \centering
    \includegraphics[width=0.47\textwidth]{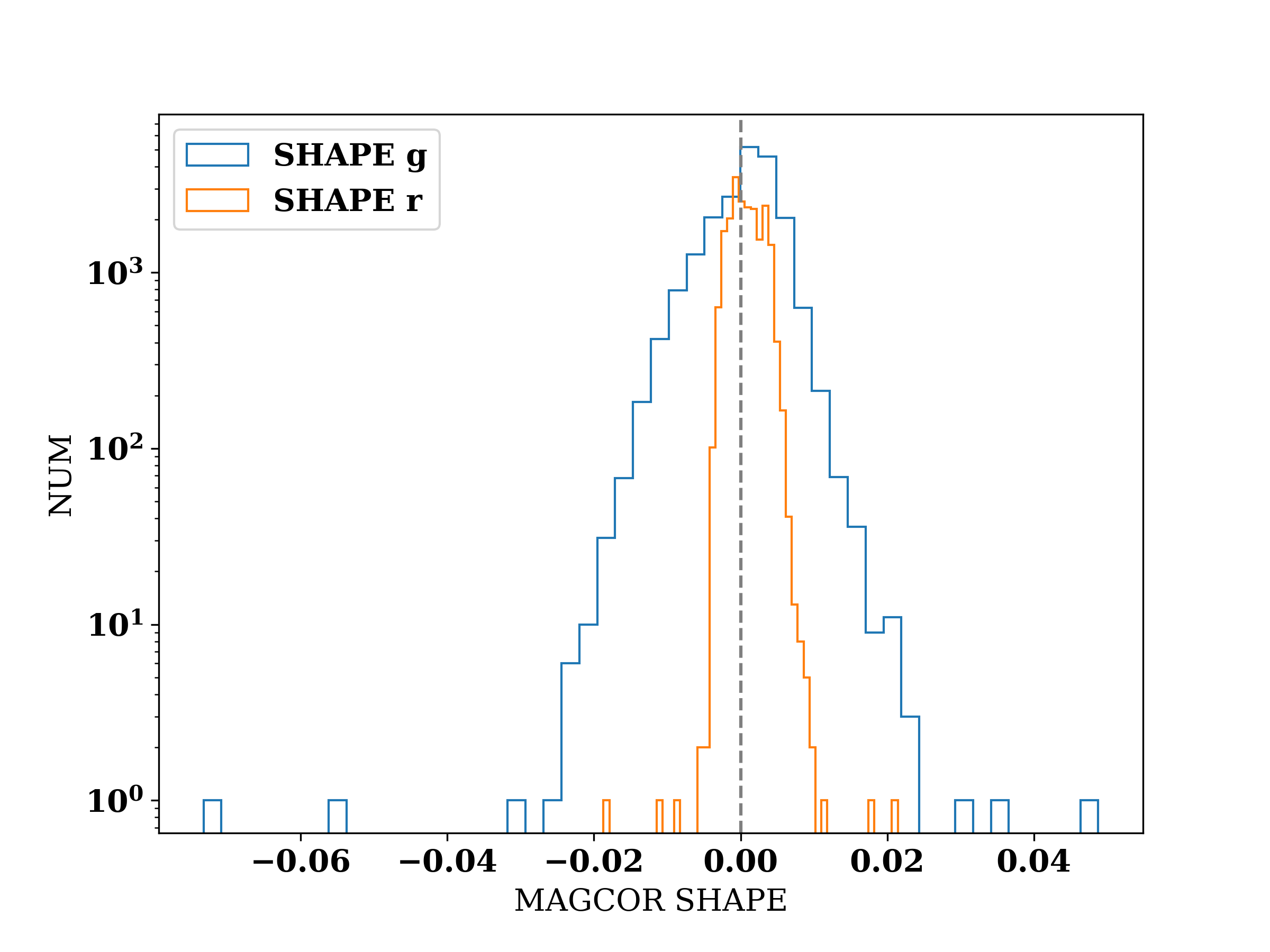}
    \caption{Magnitude corrections histogram for the \textit{SHAPE} effect; x-axis is the magnitude correction needed while y-axis is the number of observations. About 7.1\% of corrections needed are larger than 0.005 mag.}
    \label{fig:mag_corr_hist_shape}
\end{figure}

For the \textit{SHAPE} effect, we provide magnitude corrections for the \textit{g} and \textit{r} bands only, as we expect the effect to be negligible for the \textit{i} and \textit{z} bands. In Figure \ref{fig:mag_corr_hist_shape}, we see that the magnitude corrections can be both positive and negative (flux overestimated and underestimated respectively) depending on whether the SN is bluer or redder than the PSF reference star. The corrections required are noticeably larger for the \textit{g}-band. About 1.2\% of the \textit{g} and \textit{r}-band corrections are larger than 0.01 mag, while 7.1\% are larger than 0.005 mag. 

\begin{figure}
    \centering
    \includegraphics[width=0.47\textwidth]{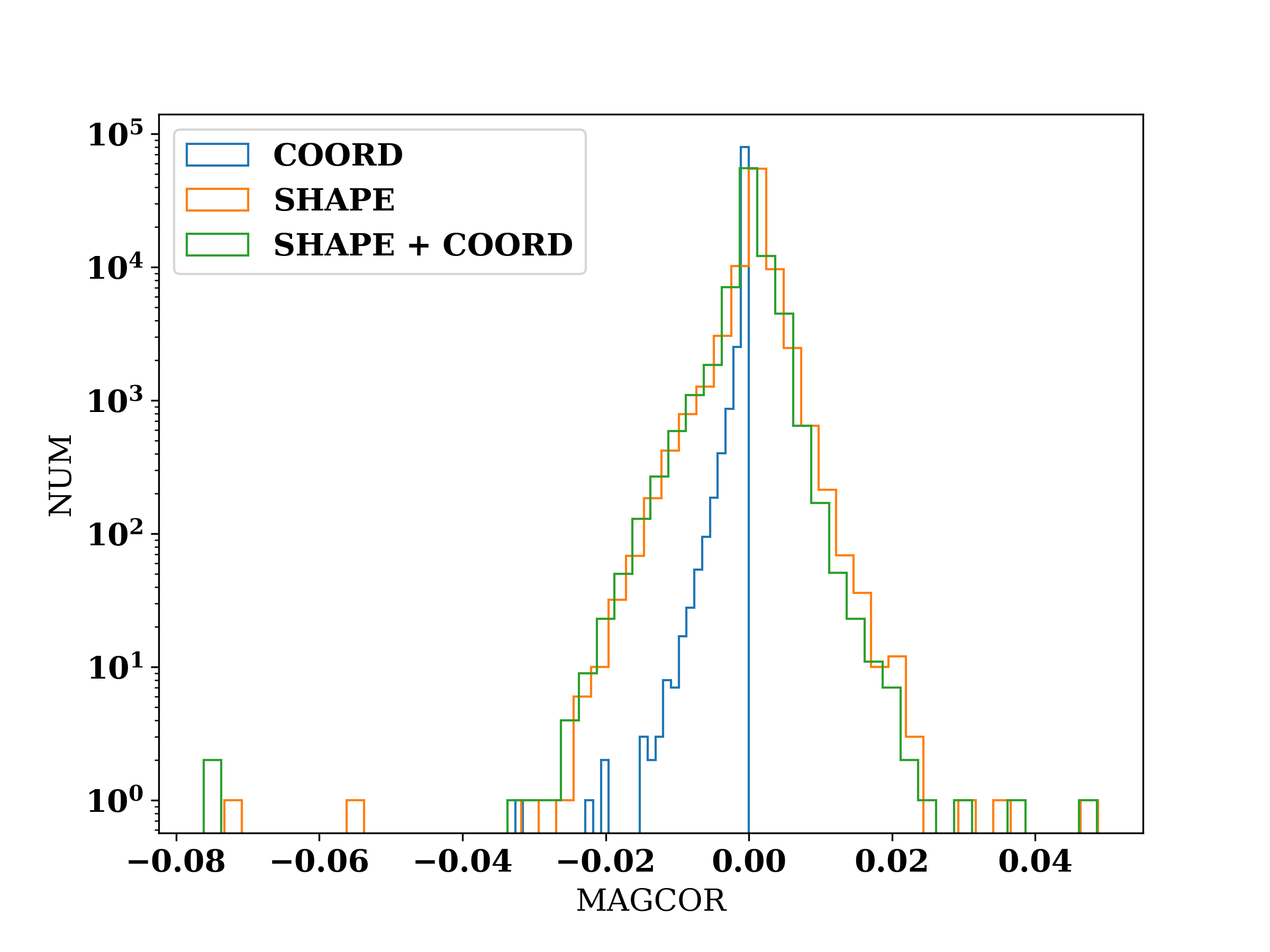}
    \caption{Magnitude corrections histogram for both \textit{COORD} and \textit{SHAPE} effects; the mean corrections required are about 3 mmag and -2 mmag for the \textit{SHAPE} and \textit{COORD} effects respectively, with the mean of the total corrections needed being about 1 mmag.}
    \label{fig:mag_corr_hist_combined}
\end{figure}

Figure \ref{fig:mag_corr_hist_combined} shows the corrections for the \textit{COORD}, \textit{SHAPE}, and both effects combined. The mean corrections required are about 3 mmag for the \textit{SHAPE} effect and -2 mmag for the \textit{COORD} effect, meaning that the mean correction is about 1 mmag when both effects are combined. 

\begin{figure*}
    \centering
    \includegraphics[width=0.98\textwidth]{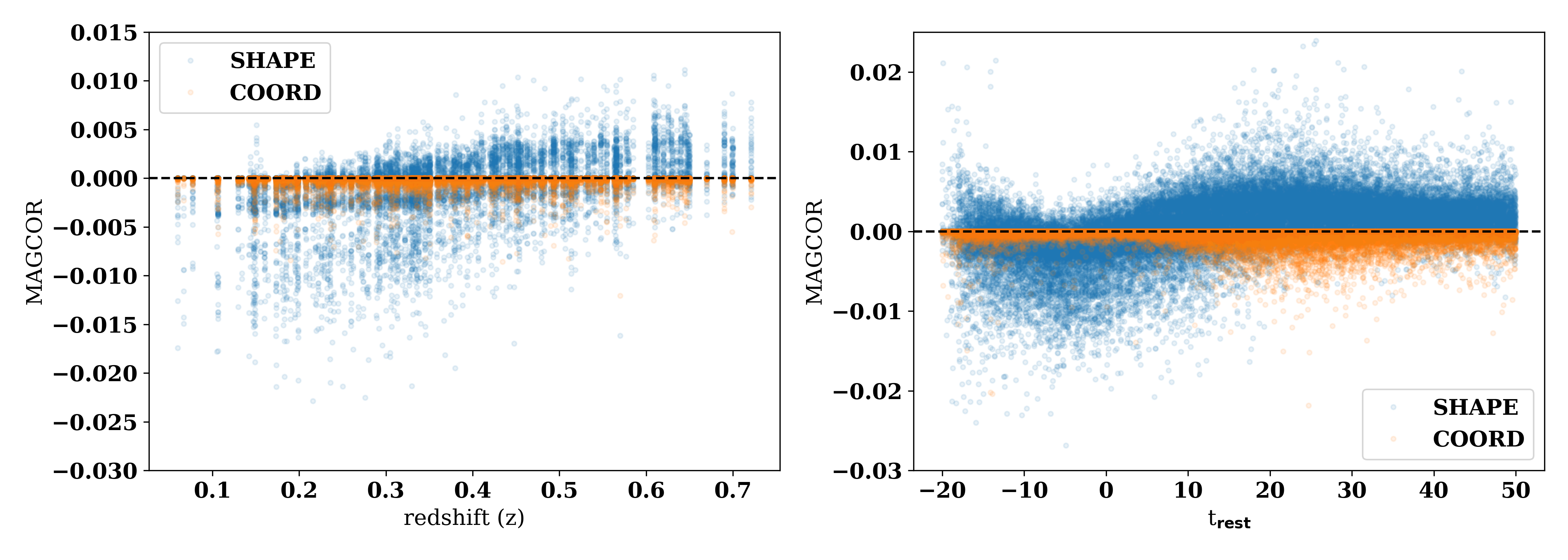}
    \caption{Magnitude corrections vs.\ redshift (left) and rest-frame epoch (right), unbinned; for the \textit{SHAPE} effect (in blue), magnitudes corrections increase as SNe are redshifted, with flux being underestimated at $z < 0.4$ and overestimated at $z > 0.5$ due to the \textit{SHAPE} effect. On the right, we also see a trend for the \textit{SHAPE} effect with SNe epoch, as the SEDs of SNe change throughout its evolution.}
    \label{fig:mag_corr_hist_z_t}
\end{figure*}

\begin{figure*}
    \centering
    \includegraphics[width=0.98\textwidth]{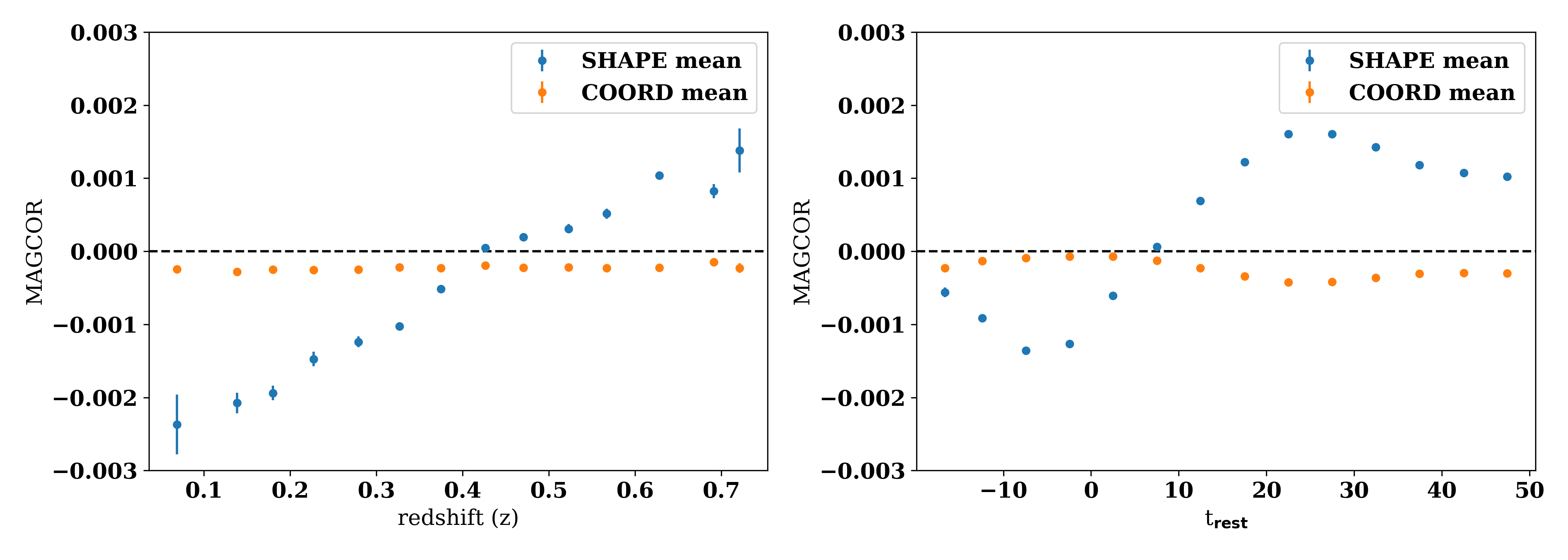}
    \caption{Magnitude corrections vs.\ redshift (left) and rest-frame epoch (right), same as Figure \ref{fig:mag_corr_hist_z_t} but binned and showing the mean and errors on the mean - the trends are much more clear for the \textit{SHAPE} effect. Note that the MAGCOR scales are different from Figure \ref{fig:mag_corr_hist_z_t}.} 
    \label{fig:mag_corr_hist_z_t_binned}
\end{figure*}

Finally, in Figure \ref{fig:mag_corr_hist_z_t} and \ref{fig:mag_corr_hist_z_t_binned} we show the dependency of the magnitude corrections on the redshift and the rest-frame epoch of the SNe, unbinned and binned. While Figure \ref{fig:mag_corr_hist_z_t} shows quite a few values where $|\text{MAGCOR}| > 0.005$, Figure~\ref{fig:mag_corr_hist_z_t_binned}, showing the mean values and errors in the mean, shows that the scatter is small except at the lowest and highest redshifts. For the \textit{SHAPE} effect, the magnitude corrections increase as the SNe are redshifted, and between $z = 0.4$ and $0.5$, where the average SN color is similar to the PSF reference star, the average magnitude correction is zero. We also see a trend for the magnitude corrections as a function of the rest-frame epoch - as the SNe becomes bluer then redder throughout its evolution, the magnitude corrections decrease then increase. 

\section{Impact on Cosmology} \label{sec:impact_cosmology}

In this section, we discuss the impact of the $\lambda$-dependent atmospheric corrections on the DES 5YR cosmological analysis.

\subsection{Distance modulus}

In this section, we discuss the effect of incorporating the \textit{COORD} and \textit{SHAPE} corrections into the DES 5YR analysis on the distance modulus $\mu$, which is given by:
\begin{equation}
    \mu = m_B + \alpha x_{1} - \beta \mathcal{C} + M_0 + \gamma G_{\mathrm{host}} + \Delta \mu_{\mathrm{bias}}
\end{equation}
where $m_B \equiv -2.5 \log (x_0)$ describes the amplitude, $x_1$ the light curve width, and $\mathcal{C}$ the color. $\alpha$ and $\beta$ describe how the SN luminosity is related to the light curve width and color respectively, while $\gamma$ describes the dependence on host-galaxy stellar mass, with $G_{\mathrm{host}} = +1/2 \quad (-1/2)$ if $M_{\mathrm{host}} > 10^{10} M_{\odot} \quad (< 10^{10} M_{\odot})$. $\Delta \mu_{\mathrm{bias}}$ is a correction for selection biases determined from simulations. In the DES-SN5YR cosmological analysis, a set of \textit{griz} light curves are fit for each SN to determine $x_0$, $x_1$ and $\mathcal{C}$ \citep{abbott2019_Y3_cosmology}. Since magnitude corrections only affect the light curve parameters ($m_B$, $\alpha x_1$, and $\beta \mathcal{C}$), we define $\mathrm{pseudo}\mathbf{\mu}$ (blinded $\mu$) using: 
\begin{multline}
    \Delta \mathrm{pseudo}\mathbf{\mu} = (m_B + \alpha x_{1} - \beta \mathcal{C})^{\mathrm{With  Corrections}} \\ - (m_B + \alpha x_{1}  - \beta \mathcal{C})^{\mathrm{Without Corrections}}    
\end{multline}
where $m_B$, $x_1$ and $\mathcal{C}$ are blinded values. Here, $\Delta \text{pseudo}\mu$ with corrections are obtained using SALT2 parameters refit after applying the $\lambda$-dependent atmospheric corrections.  We use $\alpha = 0.146$ and $\beta = 3.03$ as given in the DES-SN3YR analysis \citep{abbott2019_Y3_cosmology}.

\begin{figure*}
    \centering
    \includegraphics[width=0.98\textwidth]{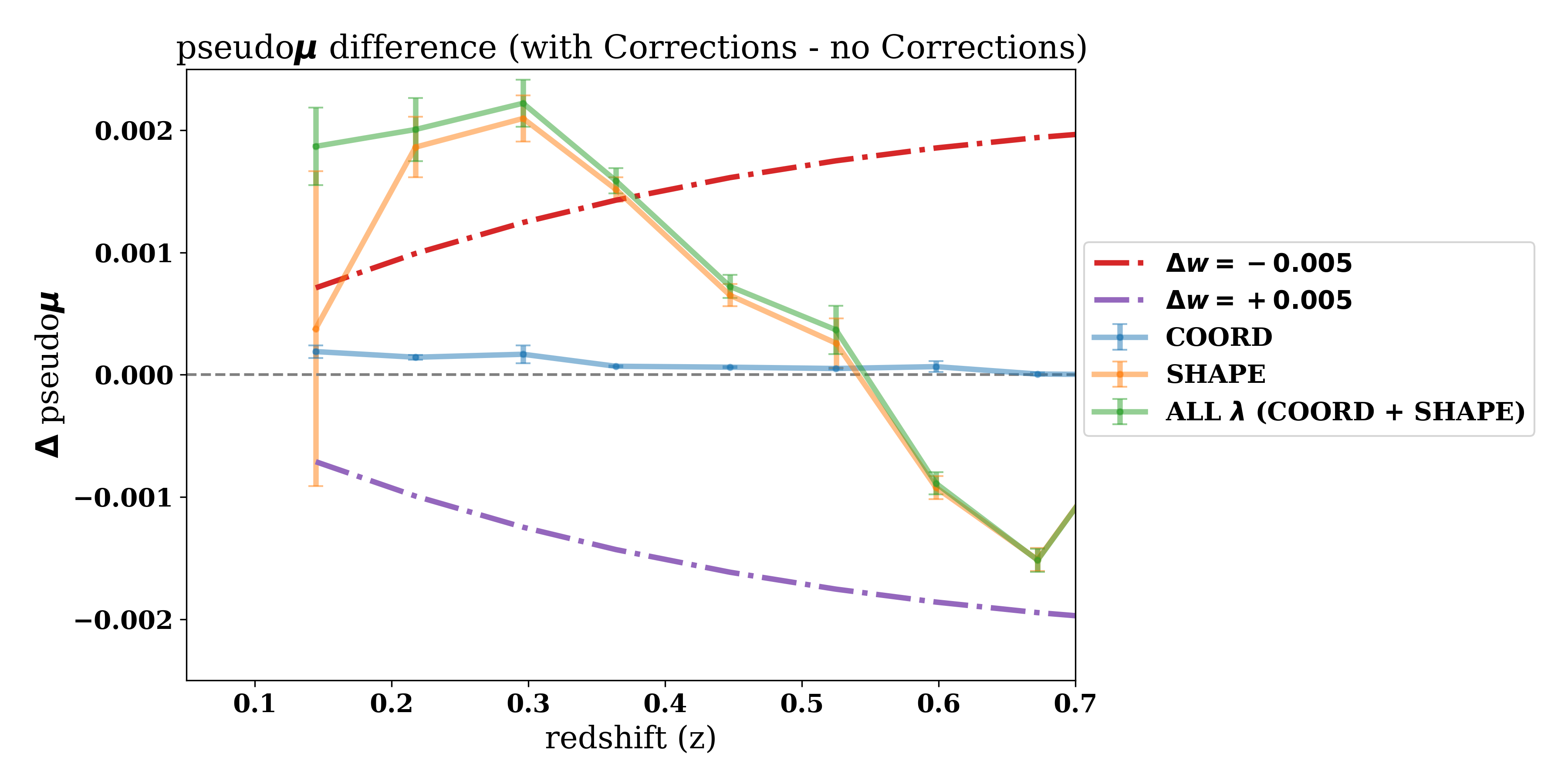}
    \caption{(Blinded) distance modulus differences (with corrections $-$ without corrections) due to $\lambda$-dependent effects; $\lambda$-dependent effects (ALL $\lambda$) shift pseudo$\mu$ by about 1 to 2 mmag at $z$ between 0.2 and 0.4 and by about $-2$ to $-1$ mmag at $z$ between 0.6 and 0.7 as the magnitude corrections for the \textit{SHAPE} effect reverse from negative to positive. A shift of $w$ by $\pm 0.005$ cause $\Delta \mathrm{pseudo} \mathbf{\mu}$ to shift by $\mp 0.0007$ to $\mp 0.002$.  At $z > 0.7$, we do not provide corrections due to the lack of \textit{g} magnitudes in that regime. }
    \label{fig:mu_w_wo}
\end{figure*}

Figure \ref{fig:mu_w_wo} shows $\Delta \mathrm{pseudo}\mathbf{\mu}$ vs.\ redshift for \textit{COORD}, \textit{SHAPE}, and all $\lambda$-dependent effects (\textit{COORD} + \textit{SHAPE}). As expected from Figure~\ref{fig:mag_corr_hist_combined}, magnitude corrections from the \textit{COORD} effect have a much smaller effect than \textit{SHAPE} corrections. However, we include both the \textit{SHAPE} and \textit{COORD} effects (ALL $\lambda$) when we discuss cosmological impacts for completeness. $\lambda$-dependent effects shift the average pseudo$\mu$ by about 1 to 2 mmag at $0.2 < z < 0.4$ and $-2$ to $-1$ mmag at $0.6 < z < 0.7$ as the magnitude corrections for the \textit{SHAPE} effect reverse from negative to positive. The $\Delta \mathrm{pseudo}\mathbf{\mu}$ values above $z>0.7$ approach zero because the majority of those SNe are not detected in the $g$ band.  Since we do not have reliable $g-i$ colors, we do not compute their magnitude corrections in $riz$, which are small compared to the $g$ band corrections. We also show in comparison the effect of a $\Delta w = \pm 0.005 $ shift on $\Delta$pseudo$\mu$ for a fiducial $\Lambda$CDM cosmology where $\Omega_m = 0.3$, $\Omega_{\Lambda} = 0.7$, $H_0 = 70$ km/s/Mpc, and $w = -1$.


\subsection{Cosmology}


We run the cosmological analysis on the (blinded) DES-SN5YR data with and without $\lambda$-dependent effects. Our results show that $w$ and $\Omega_m$ increase by less than 0.004 and 0.001 respectively, while the $1−\sigma$ statistical uncertainties are 0.02 and 0.01 respectively. We also note that $\lambda$-dependent effects were not incorporated into the low-$z$ and Foundation supernova samples, as well as at high redshifts ($z > 0.7$) where $g-i$ magnitudes are not provided, when running this analysis. Not incorporating magnitude corrections due to $\lambda$-dependent effects throughout all redshifts likely underpredict the change in $w$ and $\Omega_m$. This is evident in Figure \ref{fig:mu_w_wo}, where the slope of the ALL $\lambda$ dependent effects curve has a steeper negative slope than the $\Delta w = +0.005$ case. Since a change in slope vs.\ redshift for the distance modulus translates to a change in $w$ and $\Omega_m$, we can expect the cosmological impact to be larger than $\Delta w = +0.005$ if corrections for $\lambda$-dependent effects are applied throughout all redshifts in the DES-SN5YR data, especially at the low redshifts that anchor the distance ladder. Further cosmological analysis including only SNe at $z < 0.7$, where the $g-i$ magnitudes exist, does not show significant changes in the $w$ and $\Omega_m$ shifts, although this also reduces the constraints on these cosmological parameters. Nonetheless, we conclude that for the DES-SN5YR analysis, $\lambda$-dependent atmospheric effects are negligible compared to the statistical uncertainties.



\section{Discussion and Conclusion} \label{sec:conclusion}

In this paper, we discussed a post-processing pipeline for incorporating $\lambda$-dependent atmospheric effects on SN flux photometry for the DES-SN5YR photometry.  We divided the effects into \textit{COORD} and \textit{SHAPE}, which are biases caused by the shift of the reference star PSF position from the SN position and the differences in the shapes between the PSF and the SN respectively due to DCR and $\lambda$-dependent seeing. For the \textit{COORD} effect, we used astrometric calculations based on the data, while for the \textit{SHAPE} effect, we used image simulations using GalSIM. We found that the magnitude corrections required are -0.2 mmag and +0.3 mmag on average for the \textit{COORD} and \textit{SHAPE} effects, but roughly 0.1\% and 1\% of our measurements are larger than $\pm 10$ mmag for each of the effects respectively. The impact of $\lambda$-dependent effects on the distance modulus pseudo$\mu$ were predicted to be smaller than 0.001 for the \textit{COORD} corrections, while they range from 0.002 to -0.002 for the \textit{SHAPE} corrections depending on the redshift of the SN Ia. This amounted to shifts in $w$ and $\Omega_m$ less than 0.004 and 0.001 respectively for either effects, which are much smaller than the projected $1\sigma$ uncertainties on the data of 0.02 and 0.01 respectively. We conclude that $\lambda$-dependent atmospheric effects are not large enough to impact the DES-SN5YR analysis. 

However, future observations like LSST will have much more stringent uncertainty requirements compared to DES, and some are expected to observe in the \textit{u} band, where $\lambda$-dependent effects are much larger. In this sense, this current work provides a framework for quantifying $\lambda$-dependent effects in future work, and perhaps incorporating these effects before running SMP. 

There are, however, some limitations to our approach discussed in this paper. First, the $g-i$ magnitude may not be the best proxy for the SED of SNe. More precise DCR calculations will require using the full SED of the SN. Second, interpolating from a look-up table for the \textit{SHAPE} effect may have resulted in some errors. A way to mitigate this would be to evaluate the \textit{SHAPE} effect at the actual observed properties of the SN. Lastly, our corrections were derived by averaging over all CCDs in the exposure ignoring all focal-plane dependence on the PSF shape and astrometric solution. A more careful treatment should account for all of these effects to perform precise distance measurements with SN photometry.





\newpage

\section*{Acknowledgements}

This paper has gone through internal review by the DES collaboration. J.L. and M.S. were supported by DOE grant DE-FOA-0002424 and NSF grant AST-2108094. L.G. acknowledges financial support from the Spanish Ministerio de Ciencia e Innovaci\'on (MCIN), the Agencia Estatal de Investigaci\'on (AEI) 10.13039/501100011033, and the European Social Fund (ESF) “Investing in your future” under the 2019 Ram\'on y Cajal program RYC2019-027683-I and the PID2020-115253GA-I00 HOSTFLOWS project, from Centro Superior de Investigaciones Cient\'ificas (CSIC) under the PIE project 20215AT016, and the program Unidad de Excelencia Mar\'ia de Maeztu CEX2020-001058-M.
This research used resources of the National Energy Research Scientific Computing
Center (NERSC), a DOE Office of Science User Facility
supported by the Office of Science of the U.S. Department of
Energy under Contract No. DE-AC02-05CH11231. 

Funding for the DES Projects has been provided by the U.S. Department of Energy, the U.S. National Science Foundation, the Ministry of Science and Education of Spain, 
the Science and Technology Facilities Council of the United Kingdom, the Higher Education Funding Council for England, the National Center for Supercomputing 
Applications at the University of Illinois at Urbana-Champaign, the Kavli Institute of Cosmological Physics at the University of Chicago, 
the Center for Cosmology and Astro-Particle Physics at the Ohio State University,
the Mitchell Institute for Fundamental Physics and Astronomy at Texas A\&M University, Financiadora de Estudos e Projetos, 
Funda{\c c}{\~a}o Carlos Chagas Filho de Amparo {\`a} Pesquisa do Estado do Rio de Janeiro, Conselho Nacional de Desenvolvimento Cient{\'i}fico e Tecnol{\'o}gico and 
the Minist{\'e}rio da Ci{\^e}ncia, Tecnologia e Inova{\c c}{\~a}o, the Deutsche Forschungsgemeinschaft and the Collaborating Institutions in the Dark Energy Survey. 

The Collaborating Institutions are Argonne National Laboratory, the University of California at Santa Cruz, the University of Cambridge, Centro de Investigaciones Energ{\'e}ticas, 
Medioambientales y Tecnol{\'o}gicas-Madrid, the University of Chicago, University College London, the DES-Brazil Consortium, the University of Edinburgh, 
the Eidgen{\"o}ssische Technische Hochschule (ETH) Z{\"u}rich, 
Fermi National Accelerator Laboratory, the University of Illinois at Urbana-Champaign, the Institut de Ci{\`e}ncies de l'Espai (IEEC/CSIC), 
the Institut de F{\'i}sica d'Altes Energies, Lawrence Berkeley National Laboratory, the Ludwig-Maximilians Universit{\"a}t M{\"u}nchen and the associated Excellence Cluster Universe, 
the University of Michigan, NSF's NOIRLab, the University of Nottingham, The Ohio State University, the University of Pennsylvania, the University of Portsmouth, 
SLAC National Accelerator Laboratory, Stanford University, the University of Sussex, Texas A\&M University, and the OzDES Membership Consortium.

Based in part on observations at Cerro Tololo Inter-American Observatory at NSF's NOIRLab (NOIRLab Prop. ID 2012B-0001; PI: J. Frieman), which is managed by the Association of Universities for Research in Astronomy (AURA) under a cooperative agreement with the National Science Foundation.

The DES data management system is supported by the National Science Foundation under Grant Numbers AST-1138766 and AST-1536171.
The DES participants from Spanish institutions are partially supported by MICINN under grants ESP2017-89838, PGC2018-094773, PGC2018-102021, SEV-2016-0588, SEV-2016-0597, and MDM-2015-0509, some of which include ERDF funds from the European Union. IFAE is partially funded by the CERCA program of the Generalitat de Catalunya.
Research leading to these results has received funding from the European Research
Council under the European Union's Seventh Framework Program (FP7/2007-2013) including ERC grant agreements 240672, 291329, and 306478.
We  acknowledge support from the Brazilian Instituto Nacional de Ci\^encia
e Tecnologia (INCT) do e-Universo (CNPq grant 465376/2014-2).

This manuscript has been authored by Fermi Research Alliance, LLC under Contract No. DE-AC02-07CH11359 with the U.S. Department of Energy, Office of Science, Office of High Energy Physics.

%




\appendix

\section{Hour Angle and Air Mass} \label{sec:HA_AM}

In this section, we provide more details on the quantities relevant to DCR. 

The AM depends on the position the telescope is pointing at for a given observation and its value is usually defined to be 1.0 at the zenith. While AM is simply a function of the zenith angle (zenith to object, or the complement of altitude), we describe it in terms of the hour angle (HA) of the object in the sky as it is relevant in understanding our DCR calculations. 

\begin{figure}
    \includegraphics[width=0.55\textwidth]{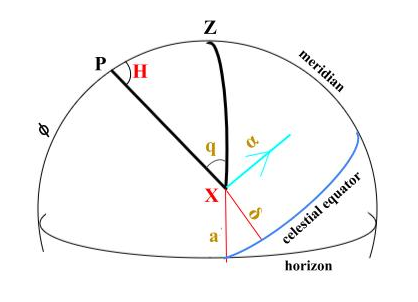}
    \centering
    \caption
    {Northern half of the celestial sphere; the HA is defined as the angle XPZ and $q$ is the parallactic angle.}
    \label{fig:celestial}
\end{figure}

Figure \ref{fig:celestial} is a diagram of the northern hemisphere of the celestial sphere, where P is the North Celestial Pole, Z is the zenith, and $\phi$ is the latitude of the observer. If X is the location of the object, $a$ is the altitude, $\alpha$ is the object’s right ascension (RA), and $\delta$ is its declination. The HA is defined as the angle XPZ. As the object traverses across the night sky (along the arrow in the diagram, parallel to the celestial equator), its HA increases. At the meridian, the HA is defined to be zero, and this is where the AM is minimum for the object on a given night. We also note that the DCR shift occurs in the direction of the altitude, and when the HA is zero, the $\delta$ and altitude directions overlap. Hence, when the HA is zero, the DCR shift only occurs in the $\delta$ direction, not the RA direction. 

\section{Astrometric Offset Decomposition} \label{sec:trig}

We decompose the astrometric offset using the spherical triangle shown in Figure \ref{fig:celestial} with the north celestial pole, the observed object, and the zenith as its vertices. The necessary angle is the parallactic angle which spans from zenith to the hour angle of a star (angle PXZ in Figure \ref{fig:celestial}). Using the spherical law of sines we find

\begin{equation}
\frac{\sin(q) }{\sin(90-\phi)} = \frac{\sin(h)}{\sin(90-\delta)},   
\label{eq:spherical_sin}
\end{equation}
 where $q$ is the parallactic angle, $\phi$ is the observer’s latitude, $h$ is the object’s hour angle, and  $\delta$ is the object’s declination. 

Equation \ref{eq:spherical_sin} is equivalent to
\begin{equation}
\sin(q) = \frac{-\cos(\phi)\sin(A)}{\cos(\delta)}
\label{eq:sin_q},
\end{equation}
which gives the sine of the parallactic angle where $A$ is the azimuth of the object.

Similarly, use of the cosine rule yields
\begin{equation}
\cos(90-a) = \cos(90-\delta)\cos(90-\phi)+\cos(\delta)\cos(\phi)\cos(h),
\label{eq:spherical_cos}
\end{equation}
where $a$ is the altitude of the object.

Equation \ref{eq:spherical_cos} is simplified to find the cosine of the parallactic angle
\begin{equation}
\cos(q) = \frac{\sin(\phi)-\sin(\delta)\sin(a)}{\cos(\delta)\cos(a)}.
\label{eq:cos_q}
\end{equation}

We then take the independent flux weighted averages of the decomposed offset derived by multiplying the sine and cosine of the parallactic angle by the interpolated offset. After removing the average for a SN candidate from every detection, we use a sum of squares to return from decomposed RA and DEC to an SMP adjusted astrometric offset.


\bibliography{refs}{}
\bibliographystyle{aasjournal}



\end{document}